\documentclass[10pt,prd,longbibliography,superscriptaddress]{revtex4-2}

\usepackage[T1]{fontenc}
\usepackage[utf8]{inputenc}
\usepackage[english]{babel}
\usepackage{gensymb}

\usepackage{amsfonts,amsbsy,amssymb,amsmath,graphicx,float} 
\usepackage{color}

\graphicspath{{figures/}}

\usepackage[linktoc=all,colorlinks=true,citecolor=blue,linkcolor=blue]{hyperref}

\begin{document}

\title{From Poincar\'e Maps to Lagrangian Descriptors: The Case of the Valley Ridge Inflection Point Potential}

\author{Rebecca Crossley}
\email{rc17354@bristol.ac.uk}
\affiliation{School of Mathematics, University of Bristol, \\ Fry Building, Woodland Road, Bristol, BS8 1UG, United Kingdom.}

\author{Makrina Agaoglou}
\email{makrina.agaoglou@bristol.ac.uk}
\affiliation{School of Mathematics, University of Bristol, \\ Fry Building, Woodland Road, Bristol, BS8 1UG, United Kingdom.}

\author{Matthaios Katsanikas}
\email{matthaios.katsanikas@bristol.ac.uk}
\affiliation{School of Mathematics, University of Bristol, \\ Fry Building, Woodland Road, Bristol, BS8 1UG, United Kingdom.}

\author{Stephen Wiggins}
\email{s.wiggins@bristol.ac.uk}
\affiliation{School of Mathematics, University of Bristol, \\ Fry Building, Woodland Road, Bristol, BS8 1UG, United Kingdom.}

\begin{abstract}
In this paper we compare the method of Lagrangian descriptors with the classical method of Poincar\'e maps for revealing the phase space structure of two degree-of-freedom Hamiltonian systems.  The comparison is carried out by considering the dynamics of a two degree-of-freedom system having a valley ridge inflection point (VRI) potential energy surface.  VRI potential energy surfaces have four critical points: a high energy saddle and a lower energy saddle separating two wells. In between the two saddle points is a valley ridge inflection point that is the point where the potential energy surface geometry changes from a valley to a ridge. The region between the two saddles forms a reaction channel and the dynamical issue of interest is how trajectories cross the high energy saddle, evolve towards the lower energy saddle, and select a particular well to enter. Lagrangian descriptors and Poincar\'e maps are compared for their ability to determine the phase space structures that govern this dynamical process.
\end{abstract}

\maketitle



\section{Introduction}
In this paper we will compare the method of Poincar\'e maps  with the method of  Lagrangian Descriptors (LDs) for the purpose of discovering and visualizing phase space structure. We begin by describing the traditional method of Poincar\'e maps and the problems that can arise in implementing and interpreting the results of this method. Then we will describe the method  of Lagrangian descriptors and discuss its features in comparison with the issues that we highlighted in our discussion of Poincar\'e maps. At the end of the introduction we will present the outline of the paper.

Poincar\'e maps have played a formative role in the development of our 
understanding of the global dynamics of two degree-of-freedom (DoF) 
Hamiltonian systems and time periodic planar dynamical systems, starting 
in the 70's and 80's, and continuing to this day (this map  is introduced 
by Poincar\'e in the third volume  of  \cite{poincare1892} and its 
properties are further developed in \cite{birkhoff1927dynamical}. The 
reader can also find a modern introduction to  Poincar\'e maps in 
\cite{contopoulos2002,wiggins2003introduction} and a discussion of its 
numerical computation   in \cite{henon1982numerical}). The development of 
the method of Poincar\'e maps occurred concurrently with the ease of 
availability of personal computing resources as this was essential for the realization and visualization of Poincar\'e maps in specific dynamical 
systems. In fact, when faced with a new and unfamiliar nonlinear dynamical system, often the first attempt at its understanding is to compute the 
Poincar\'e map. This can give a general picture of the phase space 
structure and suggest more specific and detailed avenues for analysis. 
However, the notion of a Poincar\'e map has limitations that prevent its 
applicability in many areas of current interest. These limitations are 
related to dimensionality and the difficulties in computing invariant 
manifolds of unstable periodic orbits, and we will discuss each issue  
separately.

Concerning dimensionality, Poincar\'e maps are popular for diagnosing the 
global phase space structure of two DoF Hamiltonian systems. A two DoF 
autonomous Hamiltonian system has a four dimensional phase space, but the 
dynamics are constrained to lie in the three-dimensional energy surface, 
i.e. the level set of the Hamiltonian function. The Poincar\'e map is 
constructed as follows. A two-dimensional  cross section is constructed 
transverse to the Hamiltonian flow in the three-dimensional energy 
surface. This is the domain of the Poincar\'e map (sometimes referred to 
as a Poincar\'e section, or a surface of section). The Poincar\'e map of 
this domain is constructed by associating a point in the domain with its 
first return to the domain under the action of the trajectories of 
Hamilton's equations. In this fashion, fixed points of the Poincar\'e map 
correspond to periodic orbits of the Hamiltonian system. Discussions of 
the correspondence between the phase space structure of the continuous 
time four-dimensional Hamiltonian system and the corresponding 
two-dimensional Poincar\'e map can be found in many places in the 
literature see, e.g. \cite{lichtenberg2013regular,mackay2020hamiltonian}. 
However, this approach is of limited use for visualizing and discovering 
phase space structure in Hamiltonian systems with more than two DoF. In 
particular, an $n$-DoF time-independent Hamiltonian system has a 
$2n$-dimensional phase space and a $(2n-1)$-dimensional energy surface. 
The corresponding Poincar\'e map would be $(2n-2)$-dimensional. For three 
DoF, the Poincar\'e map would be four-dimensional. Visualizing phase space structure for a 4-dimensional map unambiguously and objectively is a 
difficult task, which gets even more difficult as the number of dimensions increases. Of course, one could attempt to visualize the structures in Poincar\'e sections by considering lower dimensional slices of the Poincar\'e section. For example, in the two DoF setting, one could consider a curve in the Poincare section. However, trajectories with initial conditions starting on the curve in the two-dimensional Poincar\'e section would return to the curve with probability zero. The problem of lower dimensional sampling of Poincar\'e sections in this way becomes even more difficult with more than two degrees of freedom. We will show how the technique of Lagrangian descriptors enables us to address this problem in a new, and successful, manner.

Concerning the traditional computation of  invariant manifolds  of periodic orbits in Poincar\'e sections, we will outline the basic problems of this method. In order to compute the invariant manifolds of an unstable  periodic orbit in Poincar\'e sections we need to follow some steps. Firstly, we have to compute the periodic orbits in the system. Then we must compute the linear stability of these periodic orbits and compute the eigenvalues and eigenvectors associated with the linearization. The next step is to take many initial conditions close to the periodic orbits in the direction of the eigenvectors and to integrate them for a time  interval. In many systems the methods of finding periodic orbits cannot converge because the trajectories in the neighbourhood of the periodic orbits escape. This means that in many situations the computation of the invariant manifolds is very difficult, even using high precision. Sometimes we are not only interested in the invariant manifolds of specific periodic orbits but the invariant manifolds that exist in a specific region. In this case, the invariant manifolds can come from different periodic orbits that exist in this region. An example like this is the problem of analyzing  dynamical matching in the Caldera potential energy surface that frequently arises in the study of organic chemical reactions (see \cite{katsanikas2018phase,katsanikas2019phase,katsanikas2020a,katsanikas2020b}) where we were interested in the invariant manifolds that exist in the central area of the Caldera. 

Lagrangian descriptors have proven to be a very successful trajectory diagnostic for revealing phase space structures in dynamical systems. Originally the method was developed for analyzing Lagrangian transport phenomena in fluid dynamics \cite{chaos}, but the usefulness and applicability of the method has recently been recognized in chemistry too, see \cite{craven2016deconstructing, craven2017lagrangian, craven2015lagrangian, junginger2016transition, junginger2017chemical, feldmaier2017obtaining, patra2018detecting}. The method is very appealing since it is simple to implement computationally, the interpretation in terms of trajectory behaviour is clear, and it provides a ``high resolution'' method for exploring high-dimensional phase space with low-dimensional slices. Moreover, it applies to Hamiltonian and non-Hamiltonian systems \cite{lopesino2017} and to systems with arbitrary, even stochastic, time-dependence \cite{balibrea2016lagrangian}. Furthermore, Lagrangian descriptors can be applied to data sets directly, without the need of an explicit dynamical system \cite{mendoza2014lagrangian}.

Originally, Lagrangian descriptors were implemented computationally as follows. Each point in a chosen grid of initial conditions for trajectories in the phase space is assigned a value according to the arclength of the trajectory, starting at that initial condition that was computed for a fixed, finite time of integration, both backward and forward in time. The arclength based Lagrangian descriptor was modified in \cite{carlos,lopesino2017}, where, effectively, a different type of norm was used as a trajectory diagnostic. It has been shown that this has many advantages over the arclength. For example, it allows for a rigorous analysis of the notion of singular structures in certain cases and the relation of this notion to invariant manifolds. It also allows a natural decomposition of the Lagrangian descriptor in a way that isolates distinct dynamical effects. This was utilized in \cite{demian2017} in order to show that a Lagrangian descriptor could be used to detect the Lyapunov periodic orbits in the two degree-of-freedom Henon-Heiles Hamiltonian system.

In this original formulation of LDs, all initial conditions in the grid are integrated for the same time. This approach is known as fixed time Lagrangian descriptor. The idea is that the influence of phase space structures on trajectories will result in differences in arclength of nearby trajectories near a phase space structure. This has been quantified in terms of the notion of ``singular structures'' in the LD plots, which are easy to recognize visually, and are discussed in detail in \cite{mancho2013lagrangian,lopesino2017}. In situations where the phase space is unbounded, the fixed time LD may have difficulties resolving phase space structure as, depending on phase space structure and dynamics, many trajectories may quickly become unbounded. This has led to the development of the variable time Lagrangian descriptor (see for example \cite{katsanikas2020a,katsanikas2020b}). 

Concerning the exploration of phase space structure in high-dimensional systems, we noted earlier the difficulties in using Poincar\'e maps for this purpose, as a tightly grouped set of initial conditions may result in trajectories that become ``lost'' with respect to each other in a high-dimensional phase space. The method of Lagrangian descriptors turns this problem on its head by emphasizing the initial conditions of trajectories, rather than the precise location of their futures and pasts, after a specified amount of time. A low-dimensional ``slice'' of phase space can be selected and sampled with a grid of initial conditions at high resolution. Since the phase space structure is encoded in the initial conditions of the trajectories themselves, no resolution is lost as the trajectories evolve in time. Therefore, LDs allow us to analyze the system dynamics on low-dimensional phase space slices, since it has the capability of revealing the intersections of the invariant manifolds with these probes. 

We compare these methods  by analyzing  the dynamics of a two degree-of-freedom system having a valley ridge inflection point (VRI) potential energy surface.  VRI potential energy surfaces have four critical points: a high energy saddle and a lower energy saddle separating two wells. In between the two saddle points is a valley ridge inflection point that is the point where the potential energy surface geometry changes from a valley to a ridge. 
The region between the two saddles forms a reaction channel and the dynamical issue of interest is how trajectories cross the high energy saddle, evolve towards the lower energy saddle, and select a particular well to enter \cite{wiggins_vri, Agaoglou2020, katsanikas2020c, Victor2020}. Lagrangian descriptors and Poincar\'e maps are compared for their ability to determine the phase space structures that govern this dynamical process.

We consider the dynamics for two separate total energies. We first consider an energy below the energy of the upper energy index-1 saddle, but above the lower energy index-1 saddle. For this energy we study the transport of trajectories between the two well. Next we consider an energy above the higher energy index-1 saddle. In this case we determine the phase space structures that govern which well trajectories enter as they evolve from the high energy saddle towards the lower energy saddle. We show that the method of Lagrangian descriptors accurately reveals the relevant phase space  structures governing these processes, but require significantly less computational resources and coding complexity.

This paper is outlined as follows. In Section \ref{model}  we consider a two DoF Hamiltonian system defined by a potential energy surface having two wells, two index-1 saddles as previously studied in \cite{collins2013nonstatistical,katsanikas2020c,Agaoglou2020} in a chemical reaction dynamics context. In the Section \ref{results} we compare the analysis of the phase space structure using the traditional method of Poincar\'e maps and the method of Lagrangian descriptors. Finally, Section \ref{summary} summarizes the conclusions of this paper.

\section{Model}\label{model}
In this work we will investigate a 2 degree of Freedom (DoF) Hamiltonian model, which is the sum of the kinetic and potential energy:

\begin{equation}
\label{ham}
H(x,y,p_x,p_y)=\frac{{p_x}^2}{2 m_x}+\frac{{p_y}^2}{2 m_y}+V(x,y)
\end{equation}

\noindent
where $m_x$ and $m_y$ represent the masses of x and y respectively. 
We simplify the case we study by choosing  $m_x=m_y=1$. This can always be done by choosing mass weighted coordinates. \\

The research is focused on a potential energy surface (PES) which contains two sequential index-1 saddles and two wells: 

\begin{equation}\label{pot}
 V(x,y)=\frac{8 x^3}{3}-4x^2+\frac{y^2}{2}+xy^2(y^2-2)   
\end{equation}

\begin{figure}[htbp]
	\begin{center}
        \includegraphics[scale=0.25]{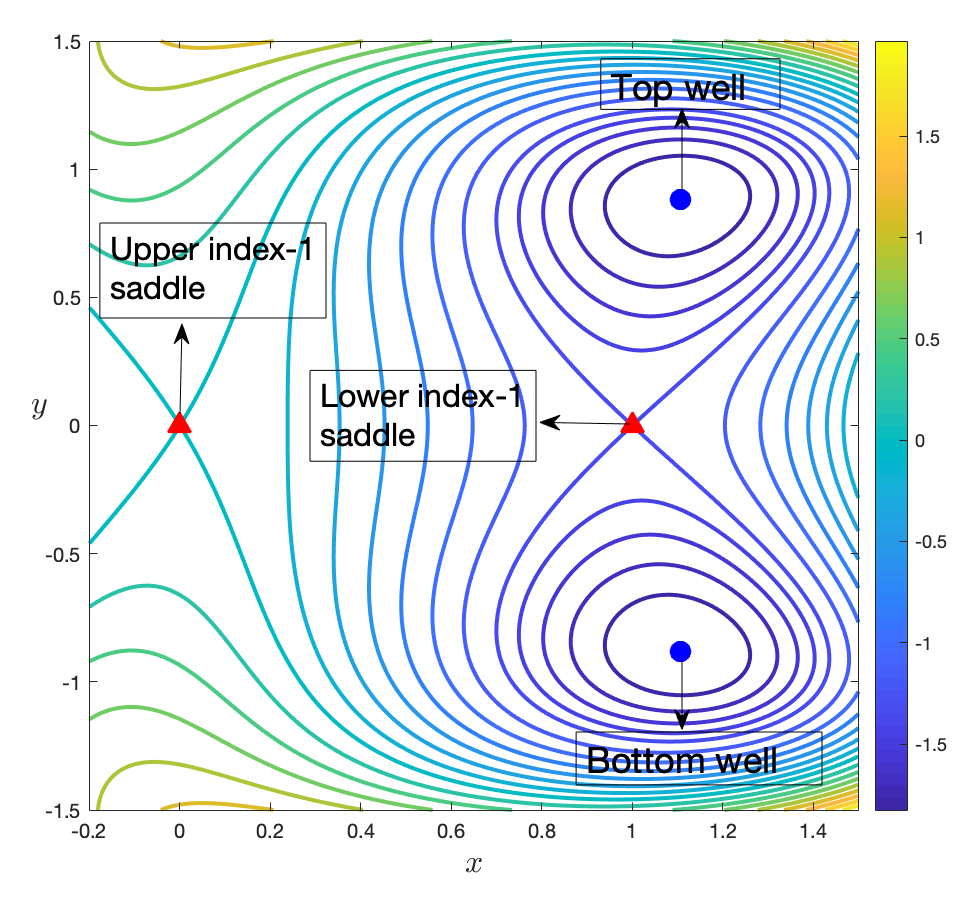} 
	\end{center}
	\caption{Equipotential contour plot of VRI. The red triangles are indicators of saddle points, where we refer to the saddle at (0,0) as the upper index-1-saddle and the other, rightmost saddle as the lower index-1-saddle. The blue circles represent the two potential wells. To distinguish between these, we use the terms top and bottom.}
	\label{VRI contours}
\end{figure}

This potential energy surface is a model for understanding certain aspects of organic chemical reactions \cite{wiggins_vri}. In this model reaction occurs when a trajectory enters the entrance region, which is that surrounding the higher energy saddle point, heading towards the lower energy saddle (see Fig. \ref{VRI contours}). 

\begin{figure}[htbp]
	\begin{center}
		\includegraphics[scale=0.3]{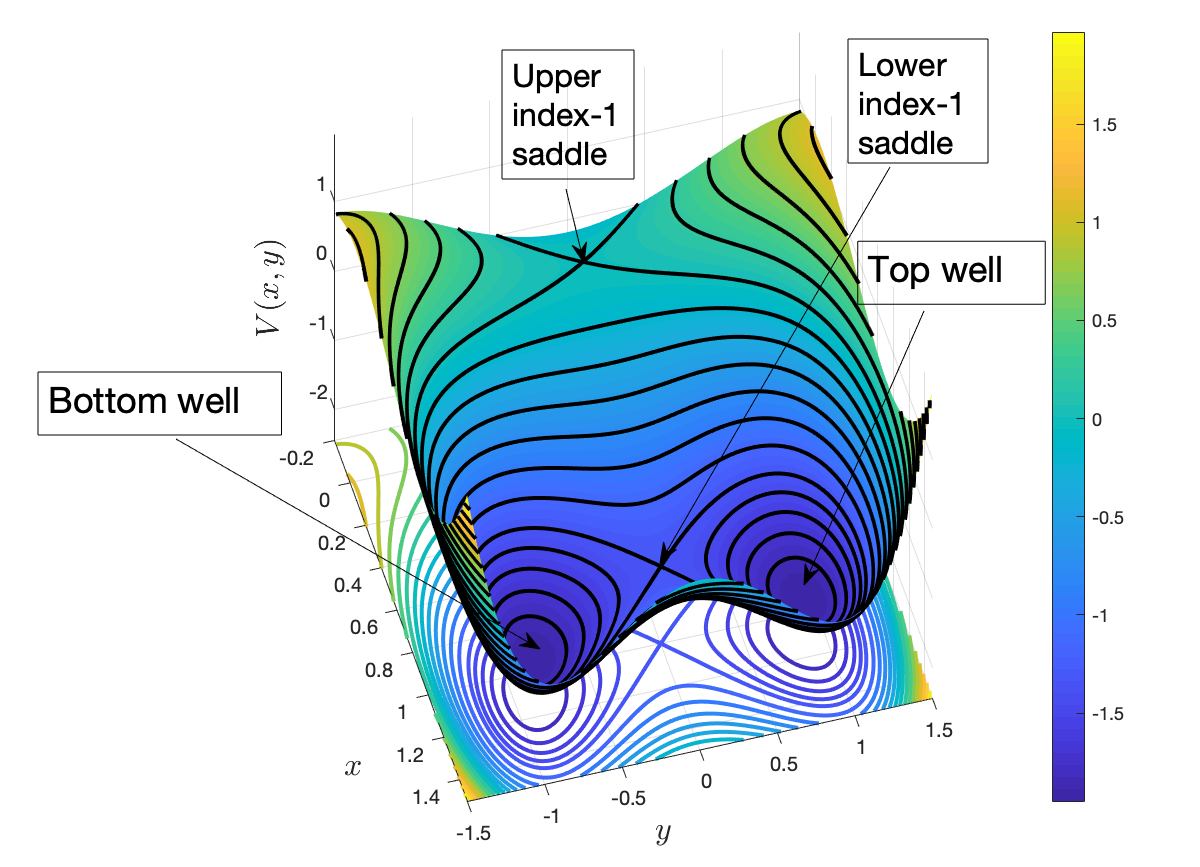} 
	\end{center}
	\caption{3D Potential Energy Surface Eq. (\ref{pot})}
	\label{3D map of VRI}
\end{figure}

We note $V(x,y)=V(x,-y)$ so the structures are symmetric under 180\degree  rotation about the origin when considering the $y-p_y$ plane (see Fig. \ref{3D map of VRI}). Hamilton's equations of motion are:

\begin{equation}
    \begin{cases}
    \dot{x}=\frac{\partial H}{\partial p_x}=p_x\\
    \dot{y}=\frac{\partial H}{\partial p_y}=p_y\\
    \dot{p_x}=-\frac{\partial H}{\partial x}=8x(1-x)+y^2(2-y^2)\\
    \dot{p_y}=-\frac{\partial H}{\partial y}=y[4x(1-y^2)-1]\\
    \end{cases}
\end{equation}

When trajectories enter  the region around the higher energy saddle with momentum directed towards the lower saddle, they begin to approach the lower energy saddle. As such, the region surrounding the higher energy saddle is referred to as the ``energy channel''. The saddle in this entrance/exit channel is referred to as the upper index-1 saddle, and it is at the origin $(0,0)$. The lower index-1 saddle at $(1,0)$ refers to the other saddle point of lower energy (RHS) which provides a sufficient energy barrier to separate the potential wells and their trajectories. We present the location and the energies of the critical points in the table \ref{tab:tab1}. 

In this work, we consider the case such that the wells are symmetric with respect to the $x$-axis. This means they are 180\degree with respect to the origin. 

If the energy is below that of the lower saddle but above the energy of the wells, then the regions of the two wells are not connected. For an overall energy level between those of the two saddles, then the entrance channel is closed and so no trajectories can escape. Some of these trajectories can move between wells via the lower index-1 saddle (see \cite{katsanikas2020c}. If the energy is above that of the upper saddle, then all forms of transport will be allowed (see  \cite{katsanikas2020c}).

\begin{table}[htbp]	
	\begin{tabular}{| l | c | c | c | c |}
		\hline
		Critical point \hspace{1cm} & \hspace{0.6cm} x \hspace{0.6cm} & \hspace{.6cm} y \hspace{.6cm} & \hspace{.2cm} \text{Potential Energy} $(V)$ \hspace{.2cm} & \hspace{.6cm} \text{Stability} \hspace{.6cm} \\
		\hline\hline
		index-1 saddle (Upper) \hspace{.5cm} & 0 & 0 & 0 & saddle $\times$ center \\
		\hline
		index-1 saddle (Lower) \hspace{.5cm} & 1 & 0 & -4/3 & saddle $\times$ center \\
		\hline
		Potential Well (Top) \hspace{.5cm} & 1.107146 & 0.879883 & -1.94773 & center \\
		\hline
		Potential Well (Bottom) \hspace{.5cm} & 1.107146 & -0.879883 & -1.94773 & center  \\
		\hline
		\end{tabular} 
		\caption{Location of the critical points of the potential energy surface.} 
	\label{tab:tab1} 
\end{table}

\section{Results}\label{results}

In this section, we will compare the methods of revealing the phase 
space structure: LDs with the traditional method of Poincar\'e maps. 
Firstly, we will compare the two methods to identify the inter-well 
transport in a system with two wells and two index-1 saddles, one of which 
is the exit/entrance channel (see Section \ref{model}). This section will be 
divided into two subsections. In the first subsection, we will describe 
the basic mechanism of the inter-well transport using the two methods 
for values of energy below the energy of the upper index-1 saddle. In the 
second subsection, we will describe the basic mechanism of the inter-well transport using the two methods for values of energy above the energy of
the upper index-1 saddle. These two subsections have a two aims. 
Firstly, we want to compare the ability of these two methods to reveal 
the phase space structures (invariant manifolds of periodic orbits) that
are responsible for the inter-well transport for energies below and 
above the energy of the upper index-1 saddle. Secondly, we want to 
compare the methods and see if one method is more effective than 
the other for detecting the homoclinic and heteroclinic intersections. 
In the first subsection we focus on the homoclinic intersections of the invariant manifolds of the unstable periodic orbits of the lower index-1 saddle because the inter-well transport is due to these kinds of intersections (see \cite{katsanikas2020c}). In the second subsection we will focus on the heteroclinic tangle of the invariant manifolds of the top and bottom unstable periodic orbits because the inter-well transport is due to this kind of tangle (see \cite{katsanikas2020c}). 

For all the computations we used (either for the classical method of Poincar\'e maps or for the method 
of the Lagrangian Descriptors) the section:

\begin{equation}
\begin{split}
\Sigma_1 & = \left\{ \left(x,y,p_x,p_y\right) \in \mathbb{R}^4 \;\Big| \; x = 1 \; , \; p_{x}\left(x,y,p_y;H_0\right) > 0 \right\} \\[.1cm]
\end{split}
\label{PSOS}
\end{equation}

\subsection{The inter-well transport below the energy of the upper index-1 saddle} 

\begin{figure}[htbp]
	\begin{center}
		\includegraphics[scale=0.6]{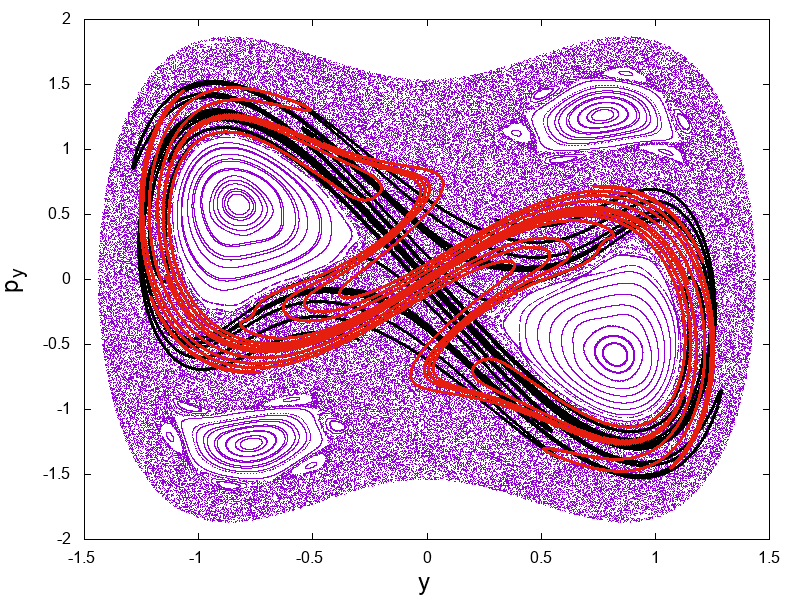}\\
        \end{center}
	\caption{The unstable (with red color) and stable invariant manifolds (with black color) of the unstable periodic orbit of the Lower index-1 saddle (that is located at the center (0,0))  and the phase space structure in the Poincar\'e section $x=1$ with $p_x>0$ for a value of Energy $E=-0.15$. The invariant manifolds are computed using the classical method.}
	\label{pos2}
\end{figure}

In this subsection we compute the invariant manifolds of the unstable periodic orbits of the lower index-1 saddle below the energy of the upper index-1 saddle. These unstable periodic orbits characterize the bottleneck region that reactive trajectories have to cross in order to evolve between wells. As an example, we  observe the unstable and stable  invariant manifolds of the unstable periodic orbit of the lower index-1 saddle in the Poincar\'e section (for  energy $E=-0.15$)  in  Fig. \ref{pos2}. These invariant manifolds are responsible for the transport of trajectories from one well to the other. This happens through  homoclinic intersections between the stable and unstable invariant manifolds of the unstable periodic orbits of the lower index-1 saddle (this is explained in detail in \cite{katsanikas2020c}). We see in  Fig. \ref{pos2} that the invariant manifolds are around the invariant curves around the stable periodic orbits of the two wells. These invariant curves represent the KAM tori (\cite{kolmogorov1954,arnold1963,moser1962}) in the Poincar\'e  section that exist around the stable periodic orbits of the two wells. Furthermore, we see two more regular regions with invariant curves on the top and bottom of the Fig. \ref{pos2}. These regular regions correspond to invariant curves around the two points of a stable periodic orbit of a family of periodic orbits with multiplicity 2 (that is associated with the two wells - see \cite{katsanikas2020c}) in the Poincar\'e section.

We first computed the invariant manifolds using classical methods and then using LDs in order to compare the two methods. We did this for a representative case (for $E=-0.15$) in the interval where the energy is lower than the energy of the upper index-1 saddle.  

\begin{figure}[htbp]
	\begin{center}
		A)\includegraphics[scale=0.30]{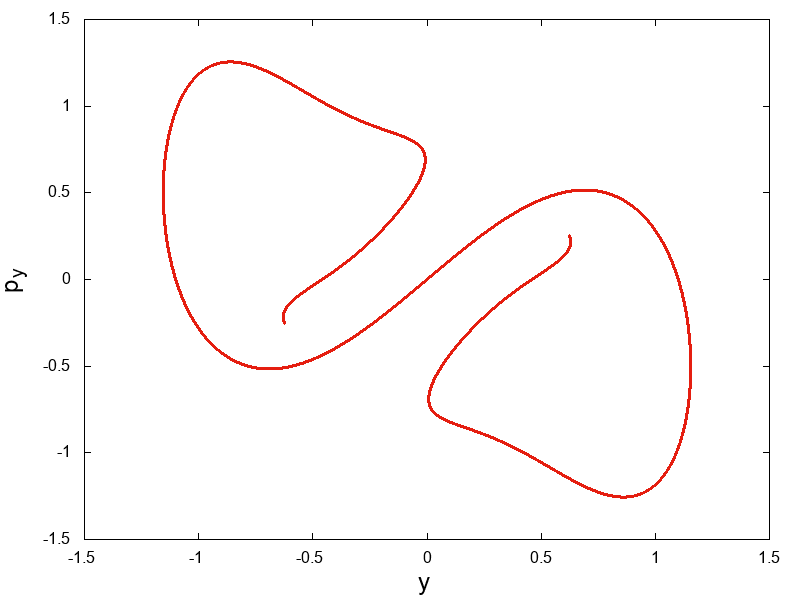}
		B)\includegraphics[scale=0.17]{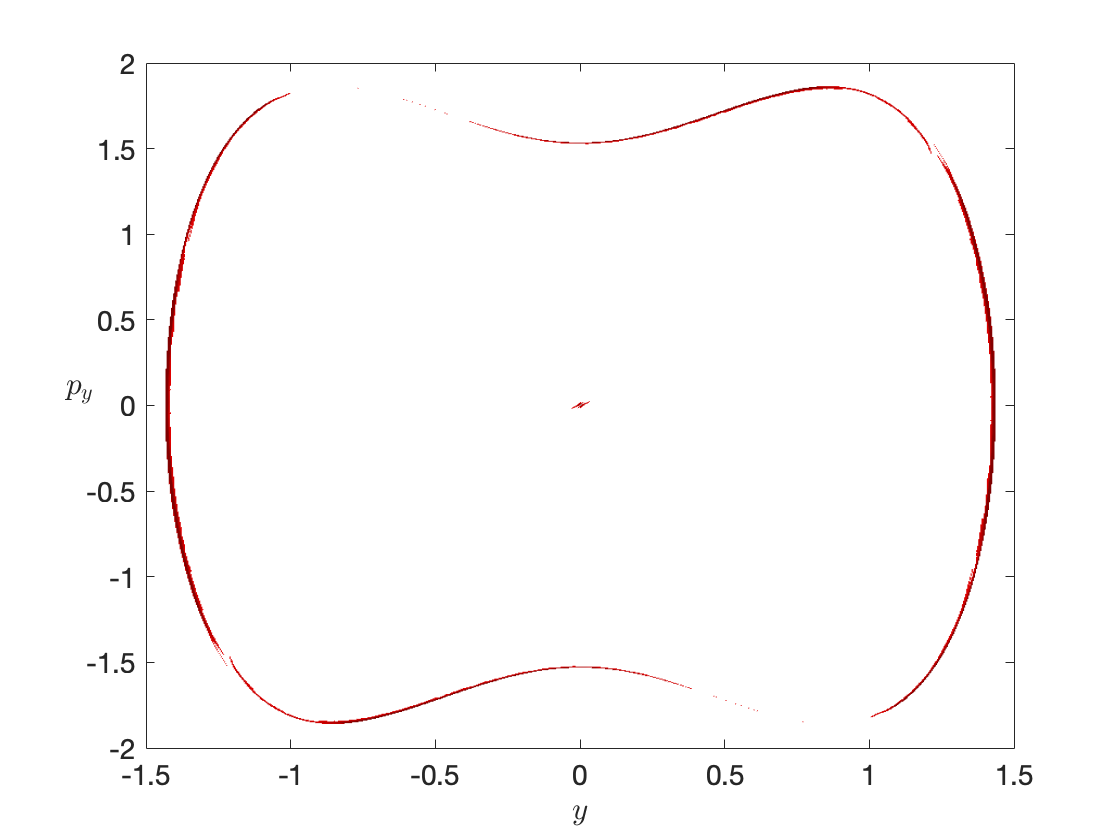}\\
		C)\includegraphics[scale=0.17]{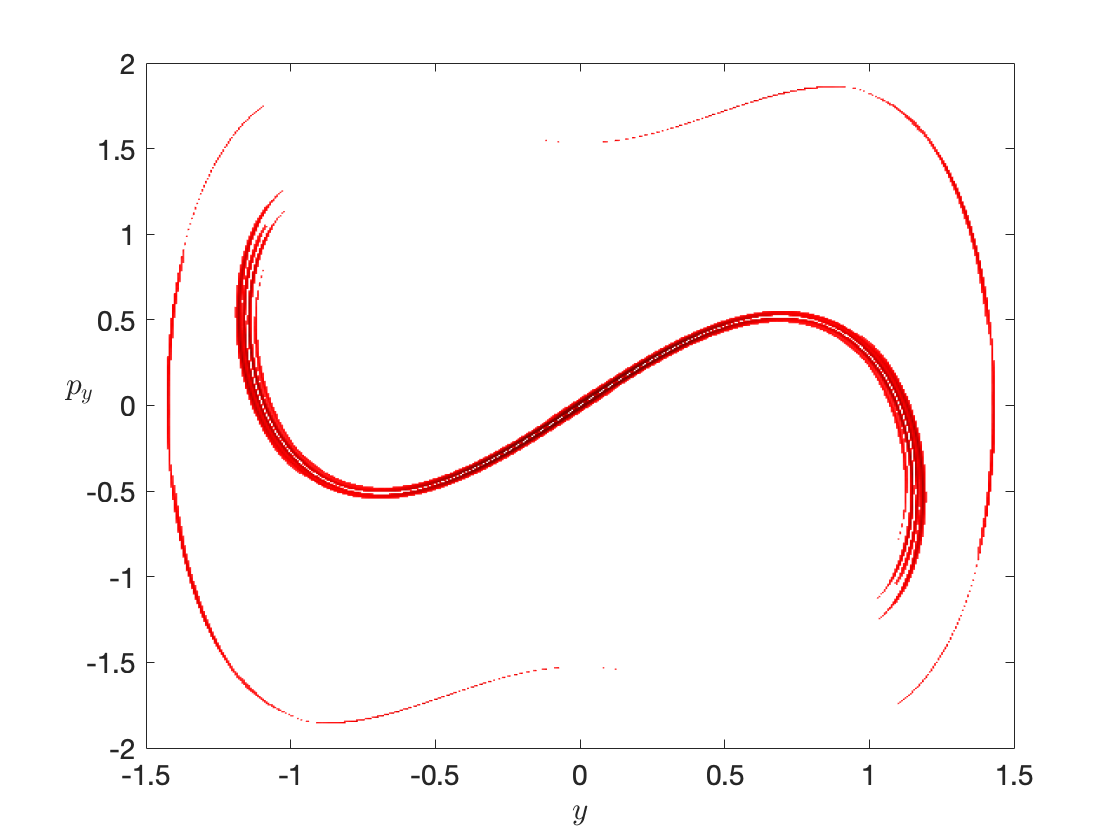}
		D)\includegraphics[scale=0.17]{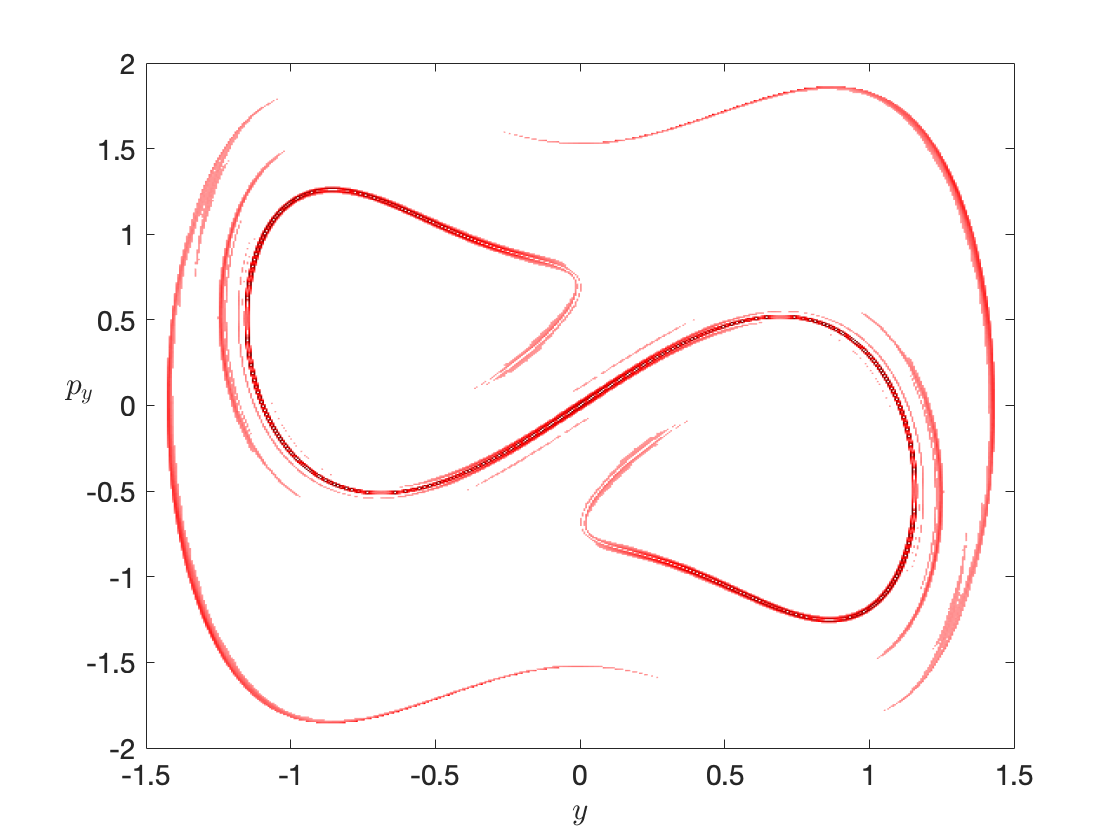}
        \end{center}
	\caption{A) The Unstable manifold of the unstable periodic orbits of the family of the lower saddle for $E=-0.15$ in the Poincar\'e section $x=1$ with $p_x>0$. This was computed using the classical method and integrating 60000 initial conditions (for 10 time units) in the direction of the unstable eigenvector. B-D) The unstable manifolds for $\tau = 2,4,6$, respectively, using the method of the LDs.}
	\label{uns1-015}
\end{figure}

\begin{figure}[htbp]
	\begin{center}
		A)\includegraphics[scale=0.3]{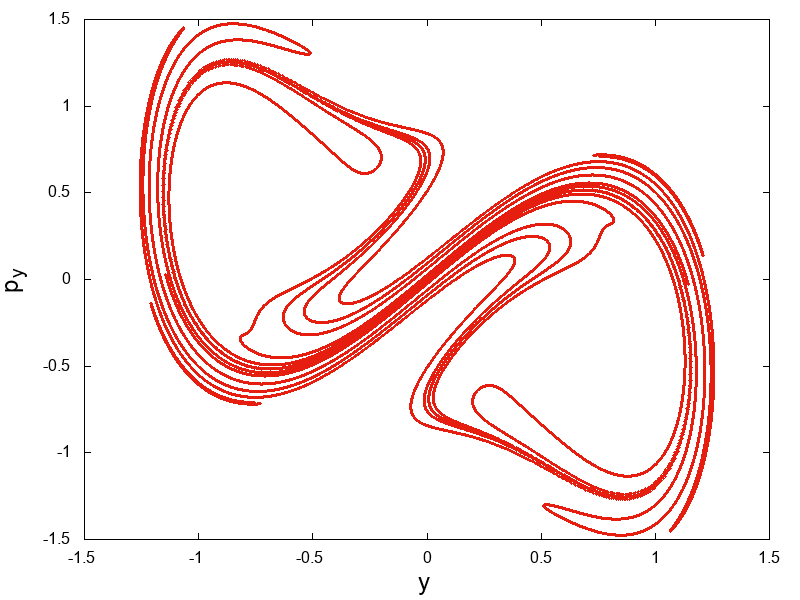}
		B)\includegraphics[scale=0.17]{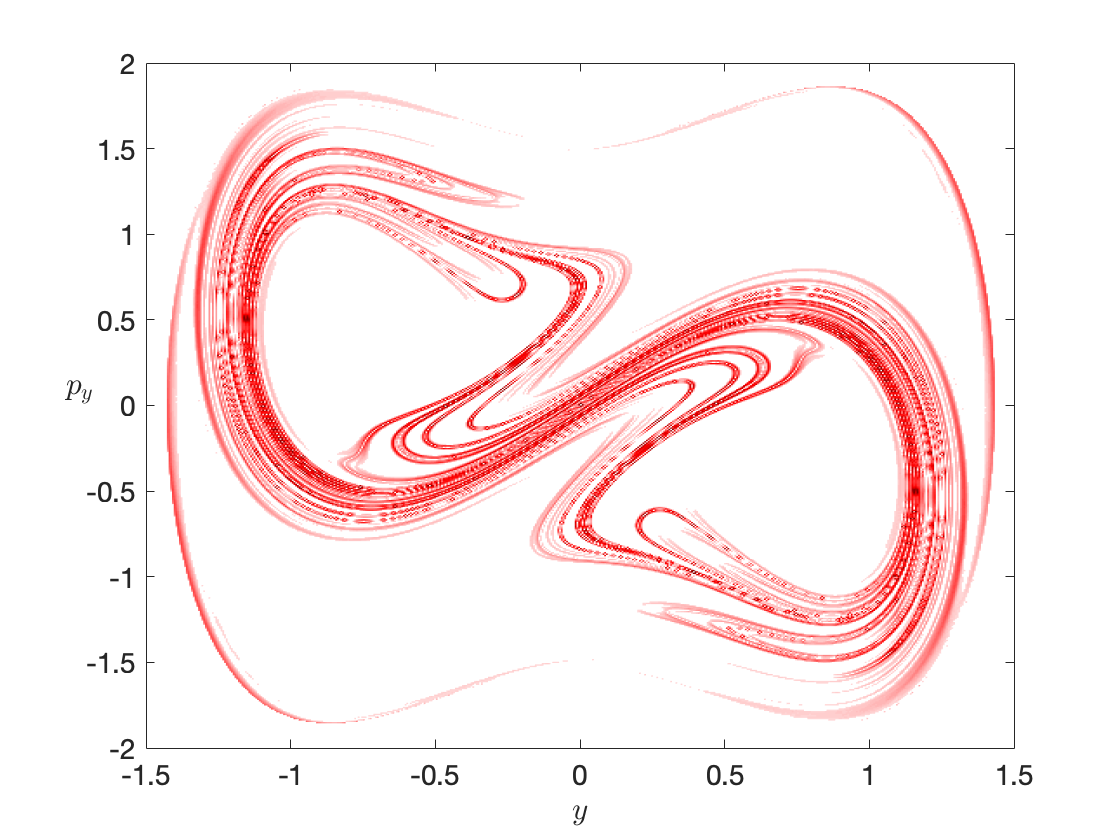}
        \end{center}
	\caption{ A) The Unstable manifold of the unstable periodic orbits of the family of the lower saddle for $E=-0.15$ in the Poincar\'e section $x=1$ with $p_x>0$. This was computed using the classical method and integrating 60000 initial conditions (for 15 time units) in the direction of the unstable eigenvector. B) The unstable manifold for $\tau = 12$ using the method of LDs.}
	\label{uns2-015}
\end{figure}

Regarding the computation of the unstable manifolds, we firstly used the classical method. As we described in the introduction we computed the periodic orbit and 
the eigenvalues and the eigenvectors in Poincar\'e  sections (for this computation see the Appendix of \cite{katsanikas2018phase}). In the case of an unstable periodic orbit (in Hamiltonian systems with two degrees of freedom) we have two eigenvalues on the real axis. The absolute value of  one of these is larger than 1 and corresponds to the unstable eigenvector. Then we integrated for a fixed time interval of 10 time units, many initial conditions (60000 in the case in this subsection) in a Poincar\'e section, close to the periodic orbit in the direction of the unstable eigenvector (panel A of Fig. \ref{uns1-015}). Afterwards, for the sake of comparison, we computed the unstable invariant manifold using the method of LDs (using the gradient of LDs - for more details see Appendix \ref{Lag Des}), for different values of the integration time $\tau$. For $\tau=2$ (panel B of Fig. \ref{uns1-015}) and $\tau = 4$ (panel C of Fig. \ref{uns1-015}) we reveal a small part of the unstable manifold that we computed using the classical methods (compare the panels A with B and C of Fig. \ref{uns1-015}). We notice that we started to approximate the unstable manifold, as computed using the classical methods in panel A, for $\tau=6$ (panel D of Fig. \ref{uns1-015}) using the method of  LDs (compare the panel A with the panel D of Fig. \ref{uns1-015}). This means that the method of the LDs begins to approximate the unstable manifold of the unstable periodic orbit of the lower index-1 saddle for $\tau=6$. Moreover in Fig. \ref{uns2-015} using the classical method we computed a larger part of the unstable manifold integrating every initial condition for 15 time units (panel A of Fig. \ref{uns2-015}). The LDs begin to approximate this unstable manifold for $\tau=12$ (compare the A with B of Fig. \ref{uns2-015}). 

\begin{figure}[htbp]
	\begin{center}
		A)\includegraphics[scale=0.30]{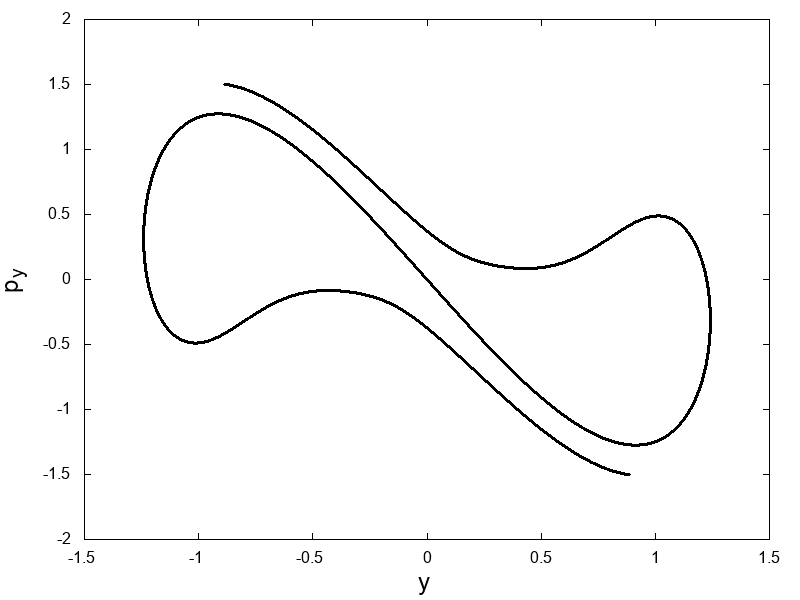}
		B)\includegraphics[scale=0.17]{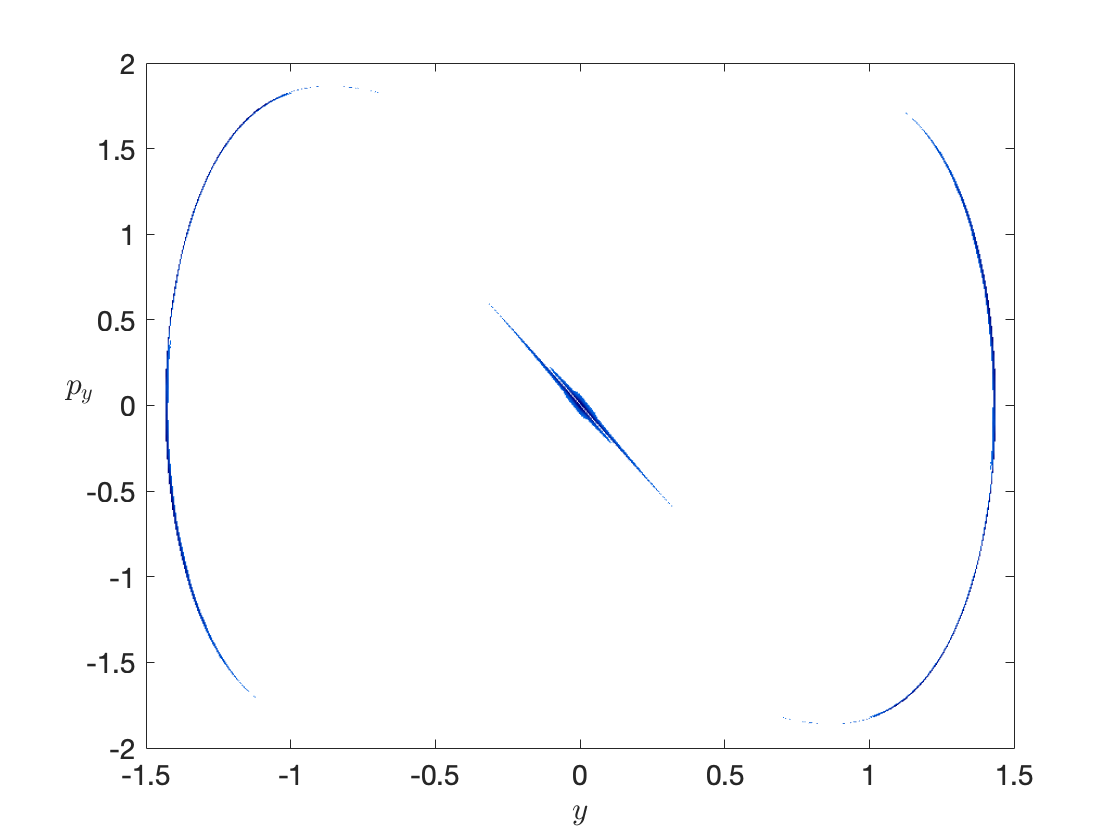}\\
		C)\includegraphics[scale=0.17]{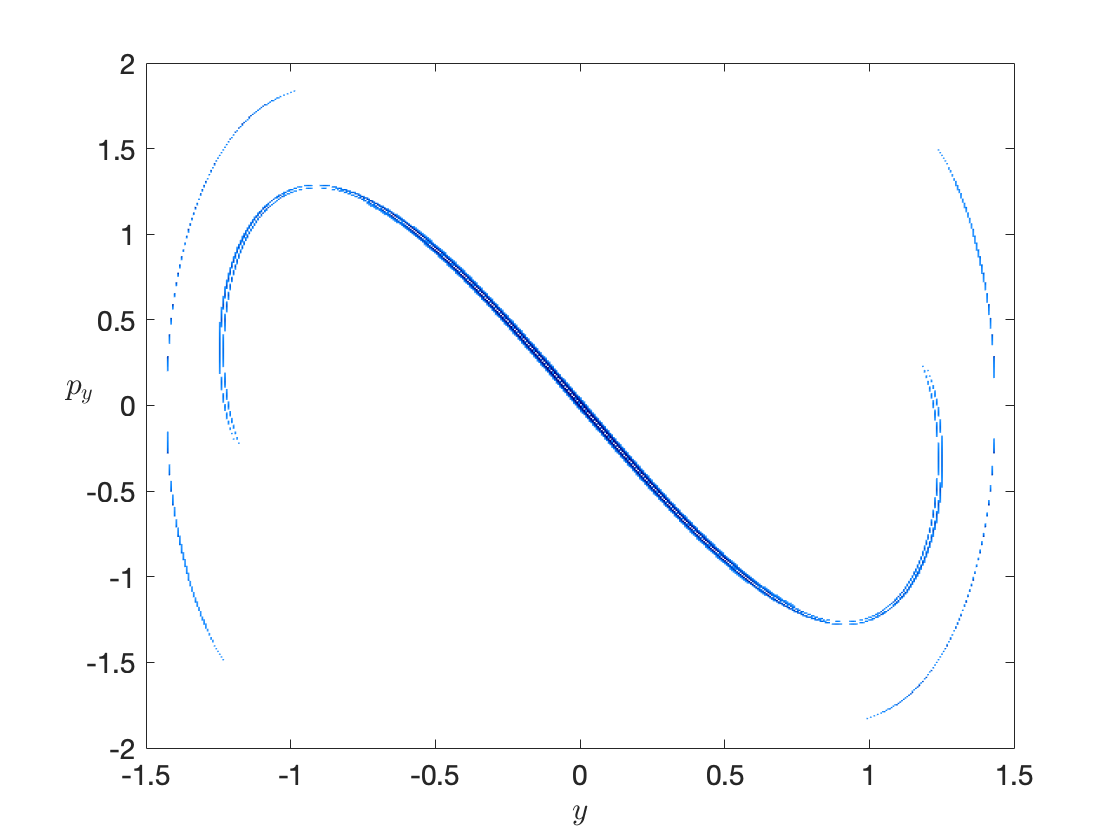}
		D)\includegraphics[scale=0.17]{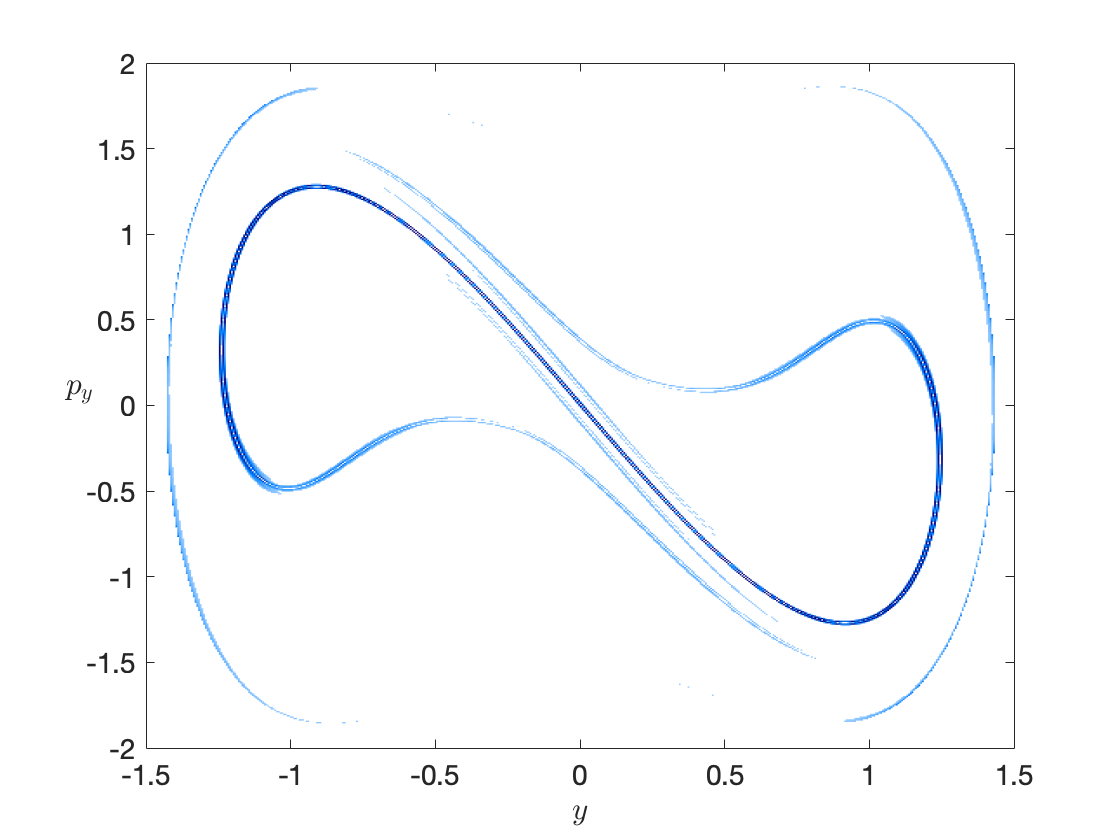}
        \end{center}
	\caption{A) The stable manifold of the unstable periodic orbits of the family of the lower saddle for $E=-0.15$ in the Poincar\'e section $x=1$ with $p_x>0$. This was computed using the classical method and integrating 60000 initial conditions (for 10 time units) in the direction of the stable eigenvector. B-D) The stable manifolds for $\tau = 2,4,6$ respectively using the method of the LDs.}
	\label{st1-015}
\end{figure}

\begin{figure}[htbp]
	\begin{center}
		A)\includegraphics[scale=0.3]{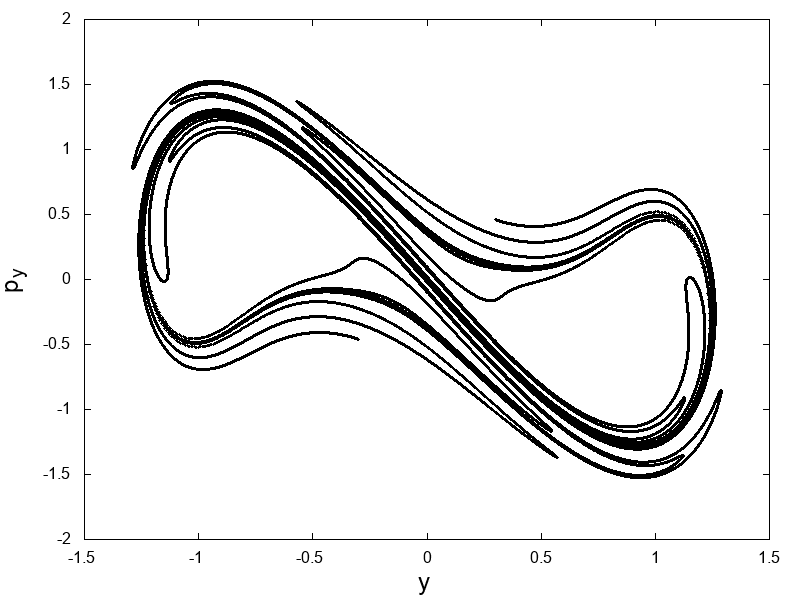}
		B)\includegraphics[scale=0.17]{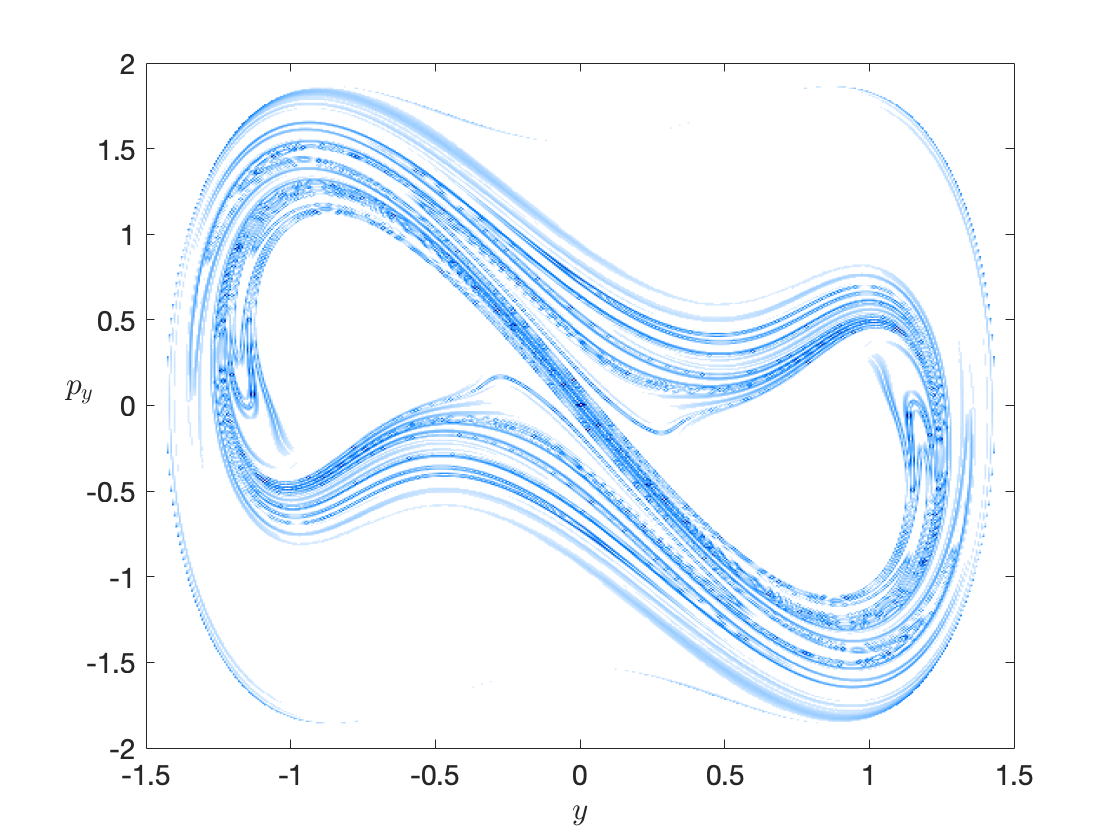}
        \end{center}
	\caption{A) The stable manifold of the unstable periodic orbits of the family of the lower saddle for $E=-0.15$ in the Poincar\'e section $x=1$ with $p_x>0$. This was computed using the classical method and integrating 60000 initial conditions (for 15 time units) in the direction of the stable eigenvector. B) The stable manifold for $\tau = 12$ using the method of LDs .}
	\label{st2-015}
\end{figure}

Regarding the computation of the stable manifolds, we initially used the classical method again (as in the case of the unstable manifolds). As we described in the introduction, for the unstable manifold we computed the periodic orbit and we computed the eigenvalues and the eigenvectors in Poincar\'e  sections (for this computation see the Appendix of \cite{katsanikas2018phase}). For the case of a stable periodic orbit (in Hamiltonian systems with two degrees of freedom) we have two eigenvalues on the real axis. The absolute value of  one of these is lower than 1 and is associated with the stable eigenvector. Then we integrated backward in time  for a fixed time interval of 10 time units many initial conditions (60000 in the case in this subsection) in a Poincar\'e section close to the periodic orbit in the direction of the stable eigenvector (panel A of Fig. \ref{st1-015}). Afterwards, for the sake of comparison, we computed the stable invariant manifold using the method of LDs (using the gradient of the LDs - for more details see Appendix \ref{Lag Des}), for different values of the integration time $\tau$. For $\tau=2$ (panel B of Fig. \ref{st1-015}) and $\tau = 4$ (panel C of Fig. \ref{st1-015})  we reveal a small part of the stable manifold that we computed using the classical methods (compare the panels A with B and C of Fig. \ref{st1-015}). We noticed that we began to approximate the stable manifold, as computed using the classical methods in panel A, for $\tau=6$ (panel D of Fig. \ref{st1-015}) using the method of the LDs (compare panel A with the panel D of Fig. \ref{st1-015}). This means that the method of  LDs begins to approximate the stable manifold of the unstable periodic orbit of the lower index-1 saddle for $\tau=6$. Moreover in Fig. \ref{st2-015} using the classical method we computed a larger part of the stable manifold integrating every initial condition for 15 time units 
(panel A of Fig.\ref{st2-015}). The LDs begin to approximate this stable manifold for $\tau=12$ (compare the A with B of Fig.\ref{st2-015}). 

\begin{figure}[htbp]
	\begin{center}
		A)\includegraphics[scale=0.30]{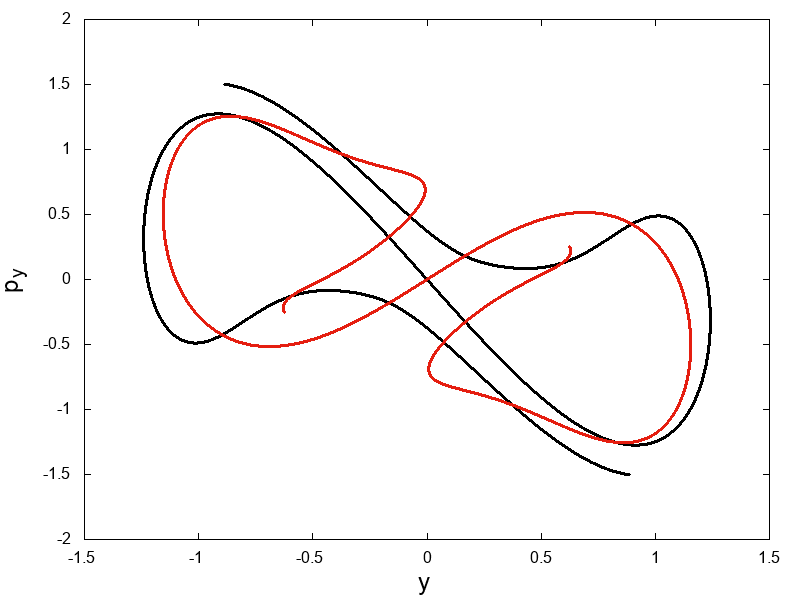}
		B)\includegraphics[scale=0.17]{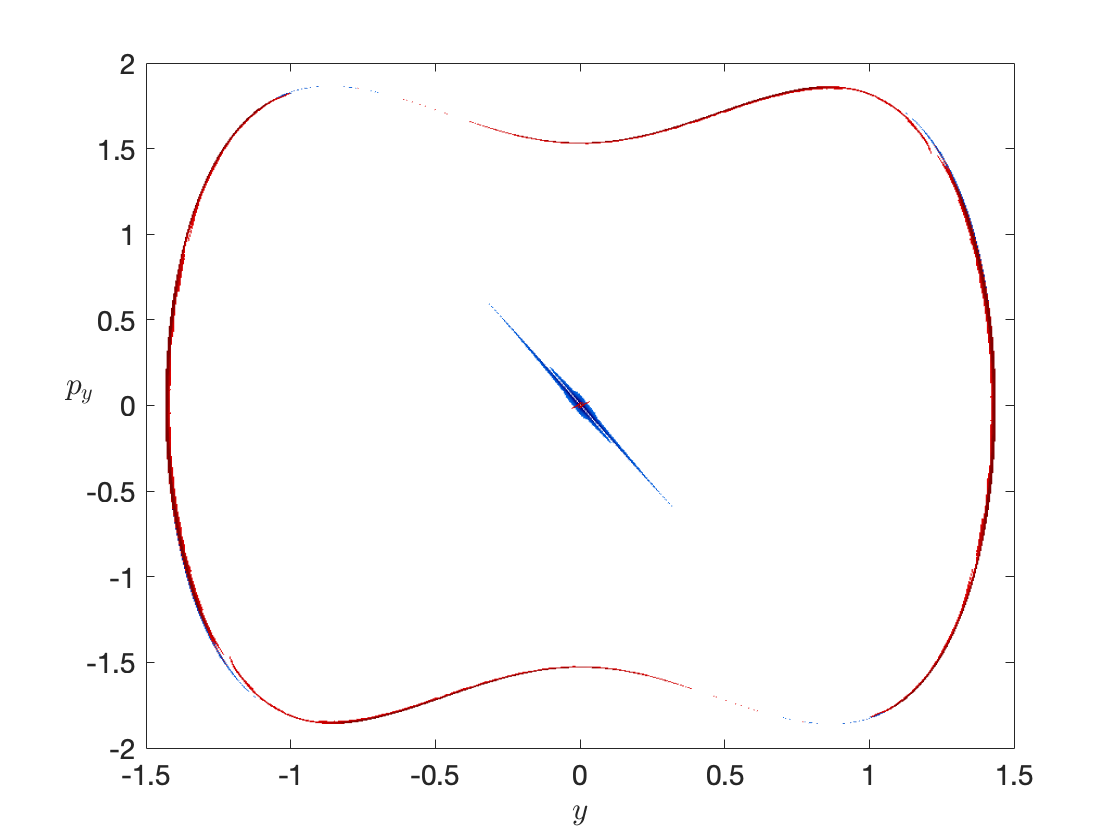}\\
		C)\includegraphics[scale=0.17]{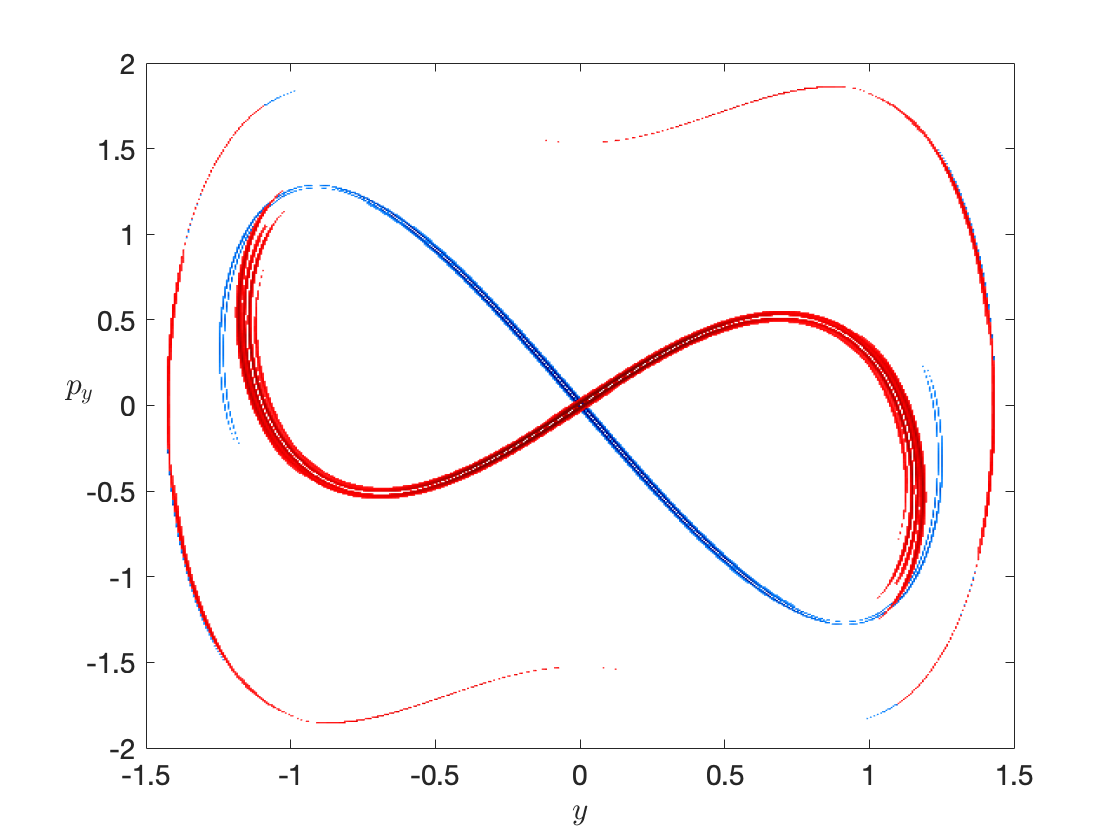} 
		D)\includegraphics[scale=0.17]{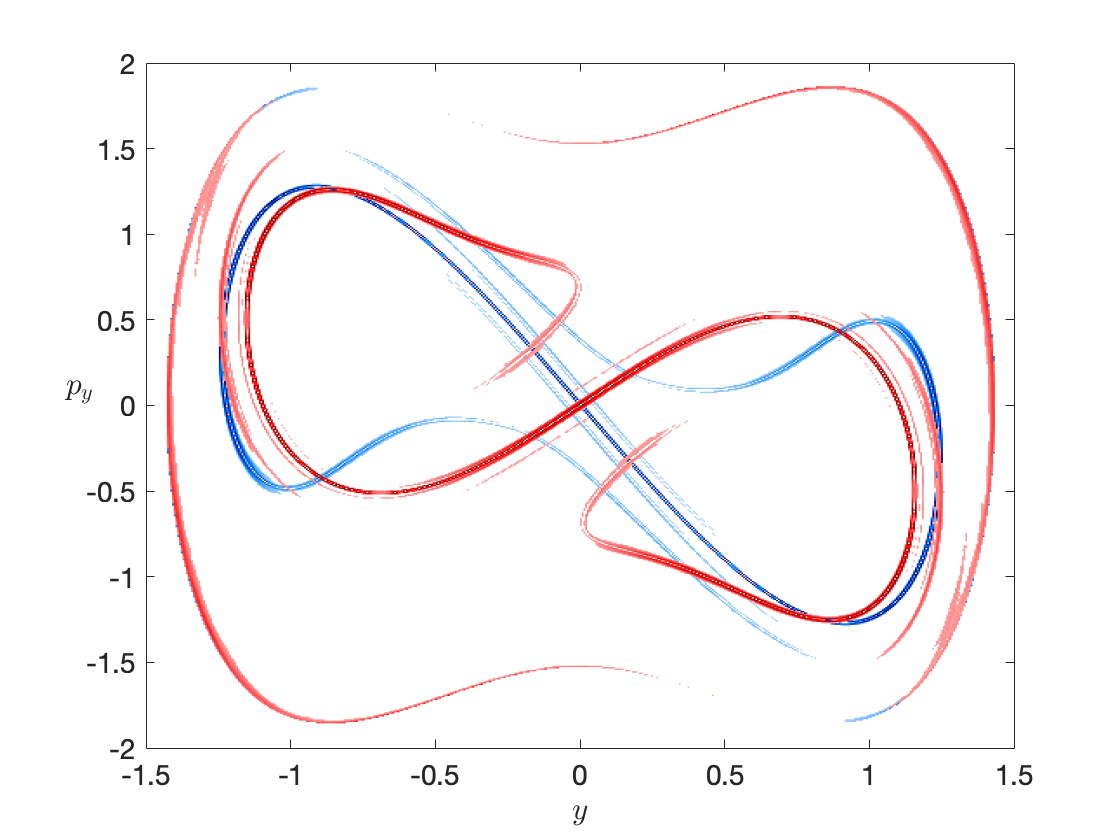}
        \end{center}
	\caption{A) The unstable (with red color) and  the stable manifold (with black color) of the unstable periodic orbits of the family of the lower saddle for $E=-0.15$ in the Poincar\'e section $x=1$ with $p_x>0$ (using the classical method). The unstable and the stable invariant manifolds  were computed using the classical method and integrating 60000 initial conditions (for 10 time units) in the direction of the unstable and stable  eigenvectors, respectively. B-D) The unstable (with red color) and the stable manifold (with blue color) for $\tau = 2,4,6$, respectively, using the method of the LDs.}
	\label{both1-015}
\end{figure}

\begin{figure}[htbp]
	\begin{center}
		A)\includegraphics[scale=0.3]{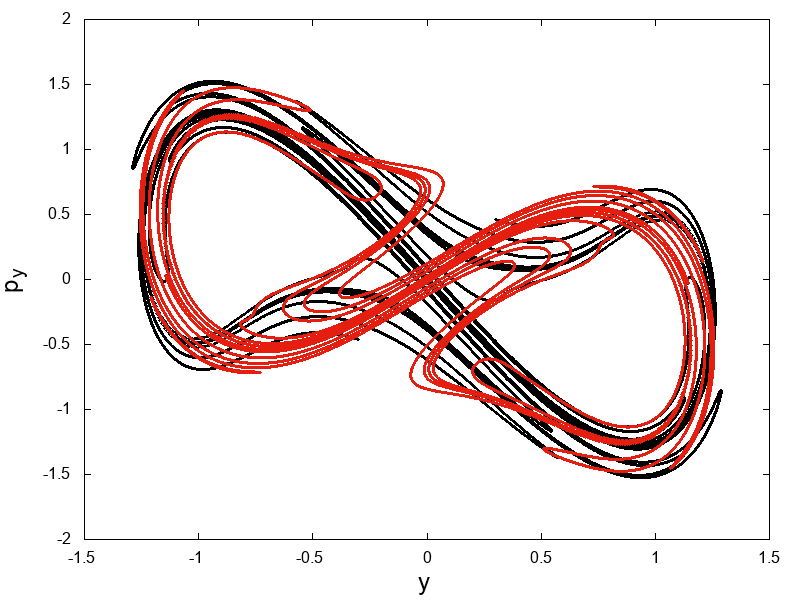}
		B)\includegraphics[scale=0.17]{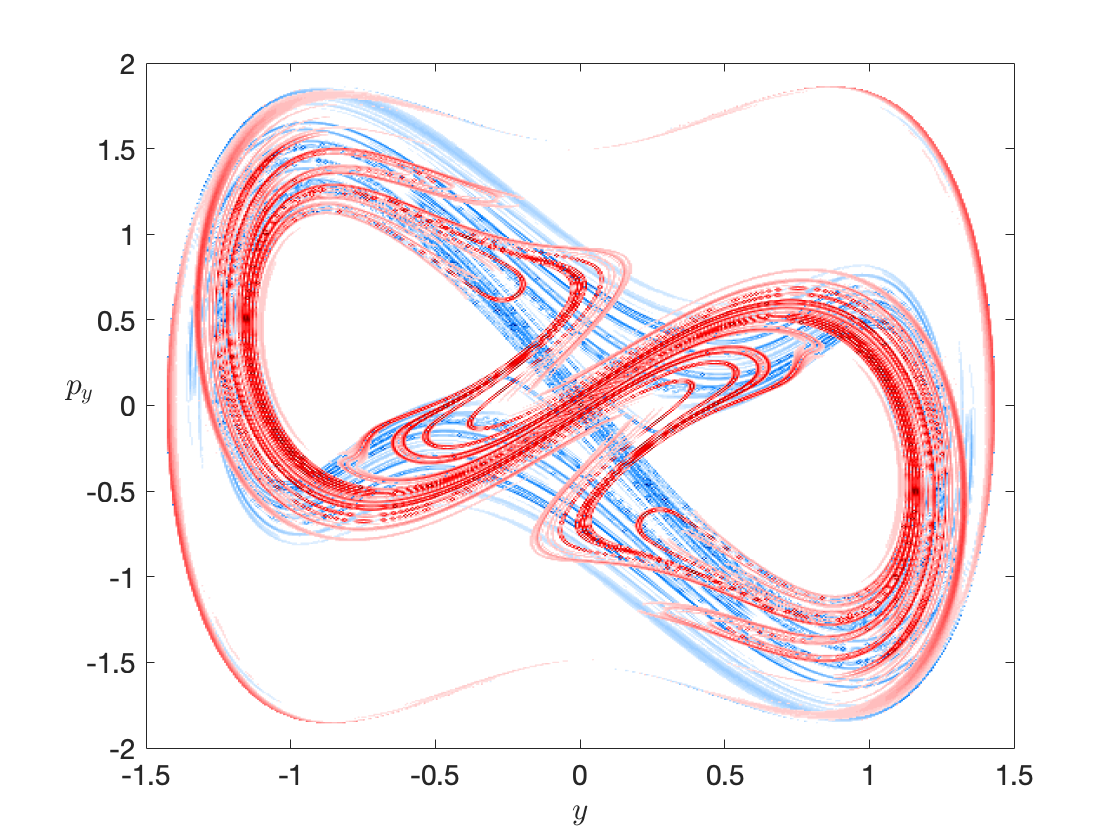}
        \end{center}
	\caption{A) The unstable (with red color) and  the stable manifold (with black color) of the unstable periodic orbits of the family of the lower saddle for $E=-0.15$ in the Poincar\'e section $x=1$ with $p_x>0$ (using the classical method). The unstable and the stable invariant manifolds  were computed using the classical method and integrating 60000 initial conditions (for 15 time units) in the direction of the unstable and stable  eigenvectors, respectively.  B) The unstable (with red colour) and the stable manifold (with blue colour) for $\tau = 12$ using the method of LDs .}
	\label{both2-015}
\end{figure}

Finally, we compare the ability of the two methods to compute the  homoclinic tangle of the invariant manifolds of the unstable periodic orbits of the lower index-1 saddle. As we can see in the panel A of Fig. \ref{both1-015} the classical method (for a time integration of 10 time units)  detects 14 homoclinic points. The LDs detect these points for integration time $\tau=6$ but fail to detect any homoclinic points for $\tau=2$ and $\tau =4$ (Fig. \ref{both1-015}). This happens because, as we explained before, the LDs begin to approximate the unstable and stable invariant manifolds for $\tau=6$. Now, if we compute a larger part of this homoclinic tangle using the classical method (for a larger integration time interval of 15 time units of the initial conditions in the directions of the eigenvectors) we see a more complicated structure as shown in the panel A of Fig. \ref{both2-015}. The LDs approximate this structure for $\tau=12$ (compare the panel A with B of Fig. \ref{both2-015}).

\subsection{The inter-well transport above the energy of the upper index-1 saddle} 

\begin{figure}[htbp]
	\begin{center}
		\includegraphics[scale=0.8]{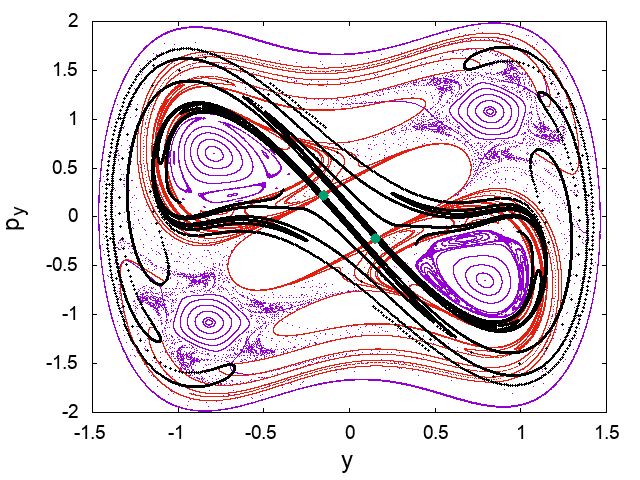}\\
        \end{center}
	\caption{The unstable (with red color) and stable invariant manifolds (with black color) of the top and bottom  unstable periodic orbits and the phase space structure in the Poincar\'e section $x=1$ with $p_x>0$ for a value of Energy $E=0.15$. We depict the top (with positive values of $y$) and bottom (with negative values of $y$) periodic orbits by green points. The invariant manifolds are computed using the classical method. }
	\label{pos2a}
\end{figure}

In this subsection we computed the invariant manifolds of the unstable periodic orbits of the top and bottom unstable periodic orbits. The family of the top and bottom unstable periodic orbits are introduced in the system after a pitchfork bifurcation of the family of unstable periodic orbits of the lower index-1 saddle just before the energy of the upper index-1 saddle and just before the family of the lower index-1 saddle vanishes. These unstable periodic orbits are important for the transport of the trajectories from one well to the other through their heteroclinic intersections (see \cite{katsanikas2020c}). This means that for example, a 
trajectory that is located in the region of the bottom well can follow the unstable manifold of the bottom unstable periodic orbit far away from this region. Then it follows, through a  heteroclinic intersection, the stable invariant manifold of the top unstable periodic orbit in the region of the top well. Actually, in this case, the invariant manifolds of the top and bottom unstable periodic orbits form a complicated heteroclinic tangle (see \cite{katsanikas2020c} and Fig. 
\ref{pos2a}). As we observed for the invariant manifolds of the unstable periodic orbits of the lower index-1 saddle (in the previous subsection), the invariant manifolds of the top and bottom unstable periodic orbits in the Poincar\'e section (for example for $E=0.15$) are around the invariant curves, that exist around the stable periodic orbits of the two 
wells (see Fig. \ref{pos2a}). We observe also the two ordered regions at the top and bottom of Fig. \ref{pos2a} (as in the previous subsection) that are characterized from the appearance of invariant curves around the points of the stable periodic orbit of the family of periodic orbits with multiplicity 2. 

\par We first computed the invariant manifolds using classical methods and then using LDs in order to compare the two methods. We did this for a representative case (for $E=0.15$) of the interval of energy above the energy of the upper index-1 saddle.

\par
\begin{figure}[htbp]
	\begin{center}
		A)\includegraphics[scale=0.39]{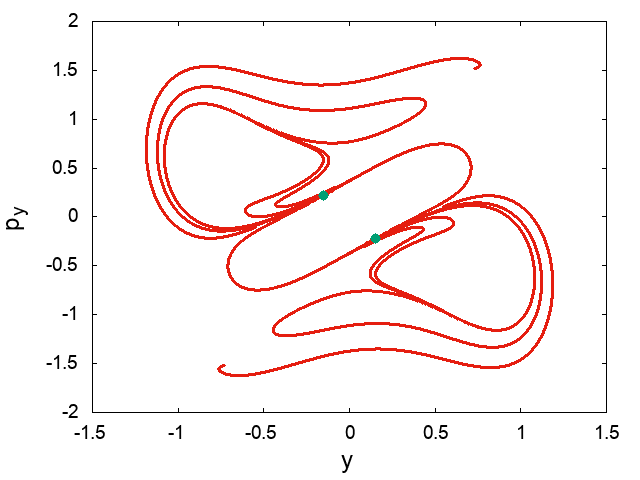}
        B)\includegraphics[scale=0.17]{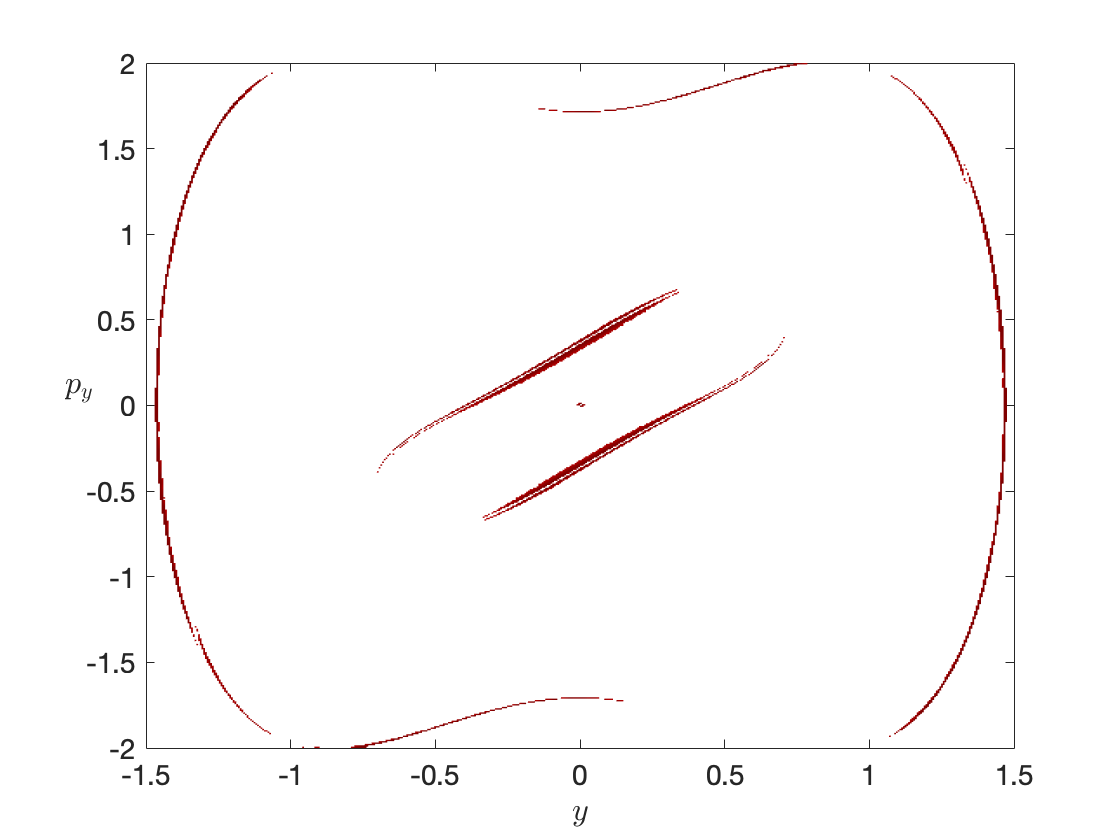}\\
        C)\includegraphics[scale=0.17]{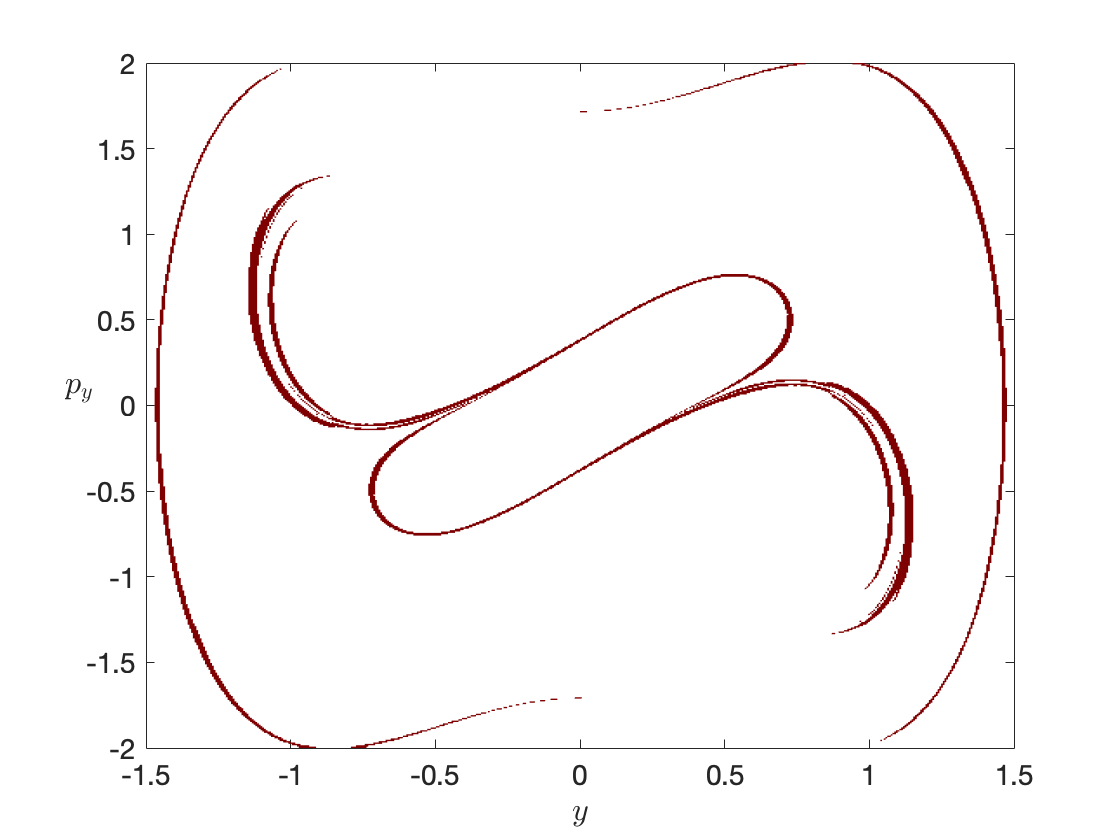}
        D)\includegraphics[scale=0.17]{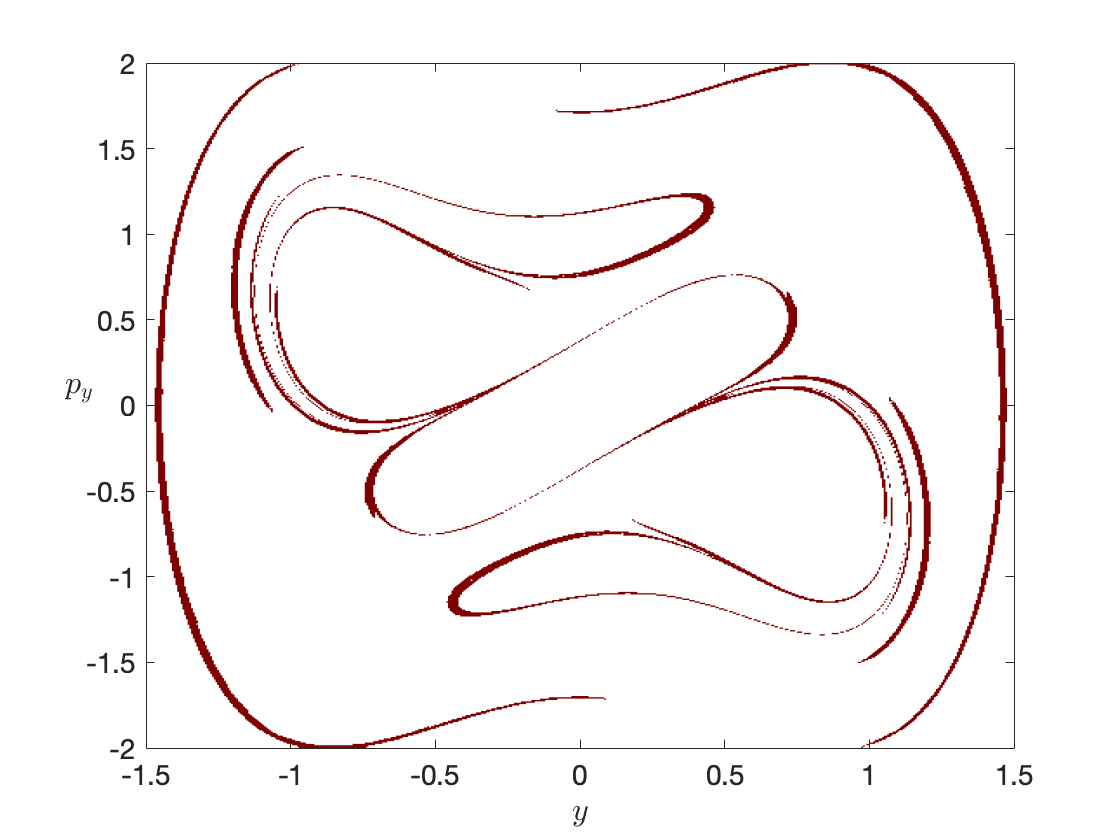}
	\end{center}
	\caption{A) The Unstable manifolds of the top and bottom unstable periodic orbits  for $E=0.15$ in the Poincar\'e section $x=1$ with $p_x>0$. This was computed using the classical method and integrating 160000 initial conditions (for 10 time units) in the direction of the unstable eigenvector. We depict the top (with positive values of $y$) and bottom (with negative values of $y$) periodic orbits by green points. B-D) Unstable manifolds for $\tau = 2,4,6$ respectively using the methods of LDs.}
	\label{uns1015}
\end{figure}

\begin{figure}[htbp]
	\begin{center}
		A)\includegraphics[scale=0.39]{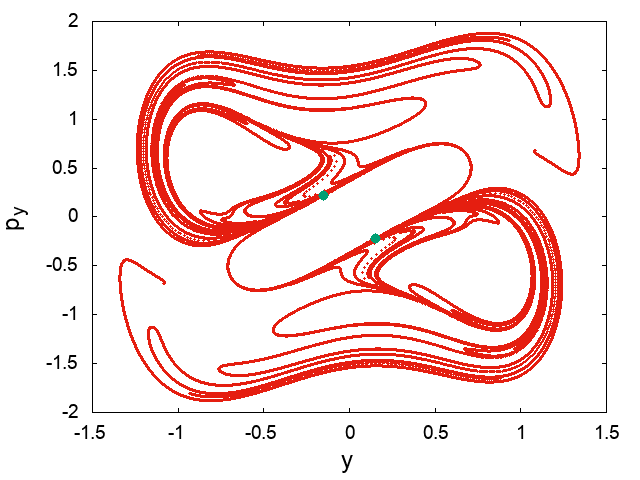}
        B)\includegraphics[scale=0.17]{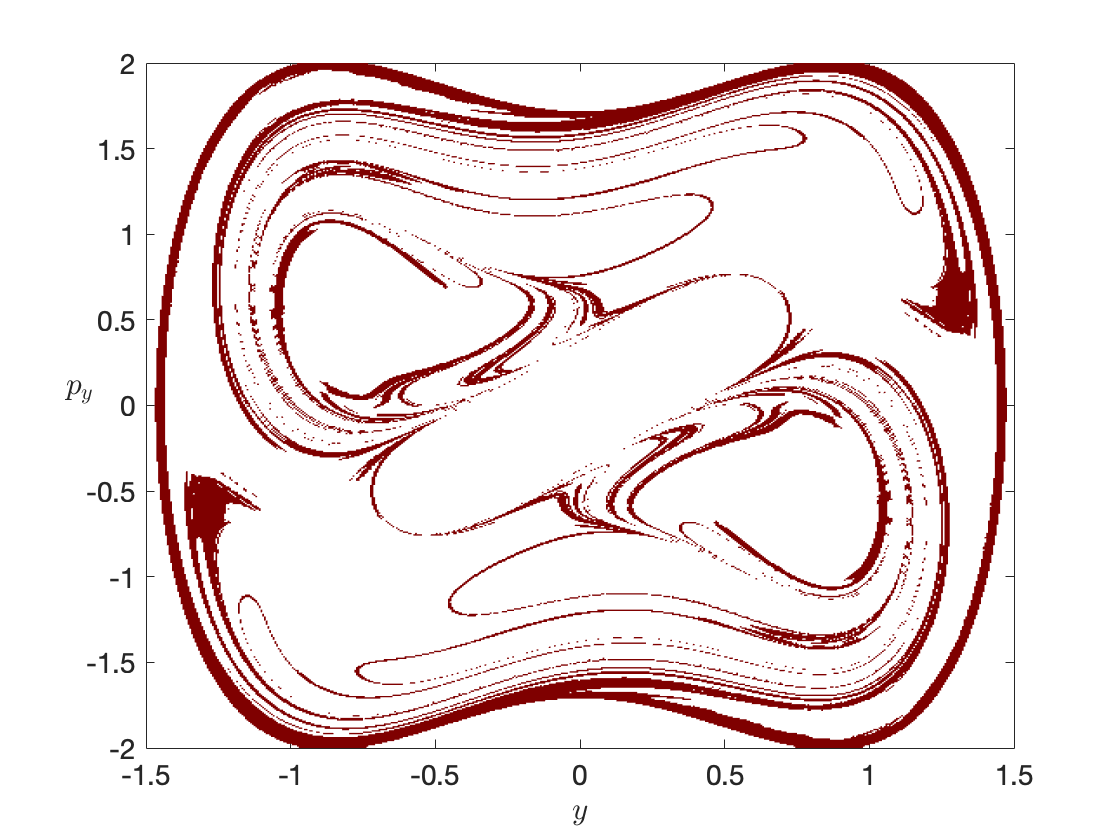}
	\end{center}
	\caption{A) The Unstable manifolds of the top and bottom unstable periodic orbits  for $E=0.15$ in the Poincar\'e section $x=1$ with $p_x>0$. This was computed using the classical method and integrating 160000 initial conditions (for 15 time units) in the direction of the unstable eigenvector. We depict the top (with positive values of $y$) and bottom (with negative values of $y$) periodic orbits by green points. B) The unstable manifolds $\tau =12$ using the methods of LDs.}
	\label{uns2015}
\end{figure}

\begin{figure}[htbp]
	\begin{center}
		A)\includegraphics[scale=0.39]{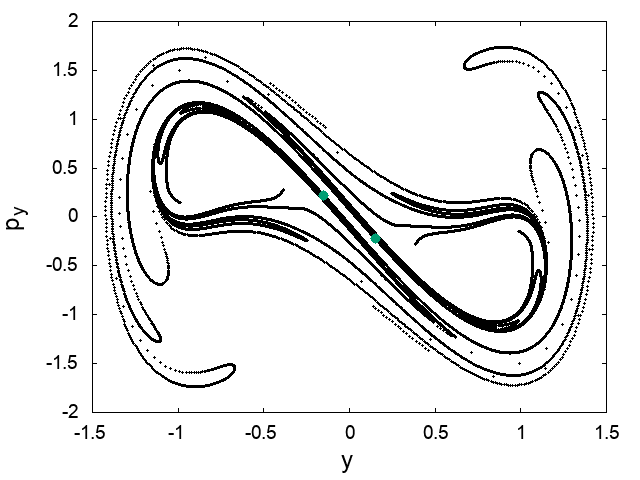}
        B)\includegraphics[scale=0.17]{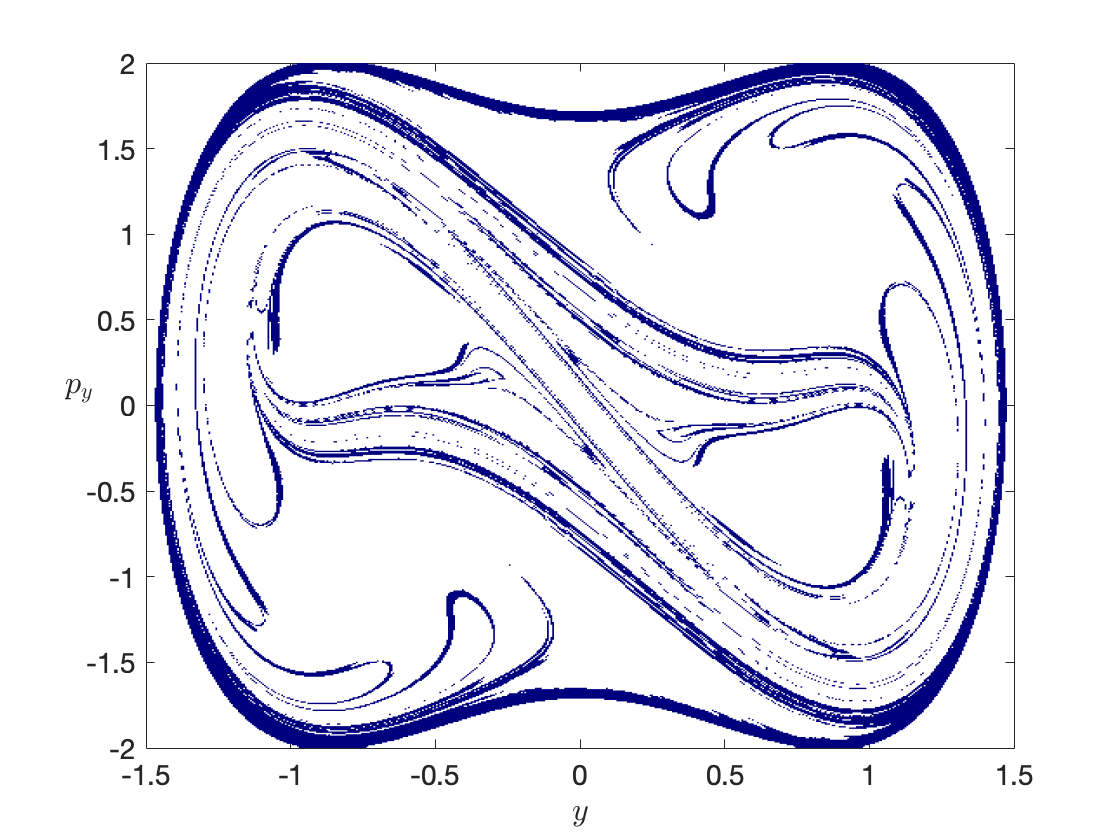}
	\end{center}
	\caption{A) The stable manifolds of the top and bottom unstable periodic orbits  for $E=0.15$ in the Poincar\'e section $x=1$ with $p_x>0$. This was computed using the classical method and integrating 160000 initial conditions (for 15 time units) in the direction of the stable eigenvector. We depict the top (with positive values of $y$) and bottom (with negative values of $y$) periodic orbits by green points. B) The stable manifolds for $\tau =12$ using the method of LDs.}
	\label{st2015}
\end{figure}

 Firstly we computed the invariant manifolds of the top and bottom unstable periodic orbits with the traditional method of Poincar\'e maps. We computed the location, the eigenvalues, and the eigenvectors of these periodic orbits. Then we integrated 160000 initial conditions forward (in the direction of the unstable eigenvectors for the unstable invariant manifolds) and backward (in the direction of the stable eigenvectors for the stable invariant manifolds) for a fixed interval of time (10 time units - see the panel A of Fig. \ref{uns1015} for the unstable manifolds and panel A of Fig. \ref{st1015}  for the stable manifold). Then we computed the LDs (through the gradient of the LDs - see Appendix \ref{Lag Des}) using different values for the integration time $\tau=2,4$ and $6$. We notice that the invariant manifolds that are extracted from the LDs start to approximate the invariant manifolds that are computed using the traditional method of Poincar\'e maps for $\tau=6$ and it fails to do this for lower values of $\tau$ (Fig. \ref{uns1015} for the unstable manifolds and Fig. \ref{st1015} for the stable manifolds). Now if we use the classical method of Poincar\'e  maps to compute a larger part of the invariant manifolds (for a time interval of 15 time units -  panel A of Fig. \ref{uns2015} for the unstable manifolds and panel A of Fig. \ref{st2015}  for the stable manifold) we find more complicated structures. These structures are approximated from the corresponding invariant manifolds that are extracted from the LDs for $\tau=12$
(Fig. \ref{uns2015} for the unstable manifolds and Fig.\ref{st2015} for the stable manifolds).

\begin{figure}[htbp]
	\begin{center}
		A)\includegraphics[scale=0.39]{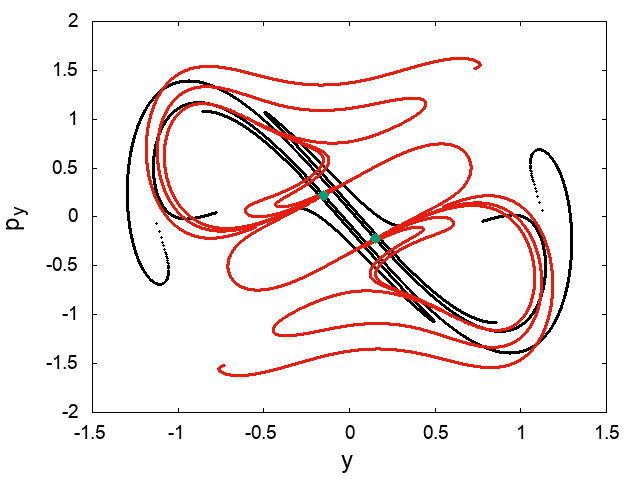}
        B)\includegraphics[scale=0.17]{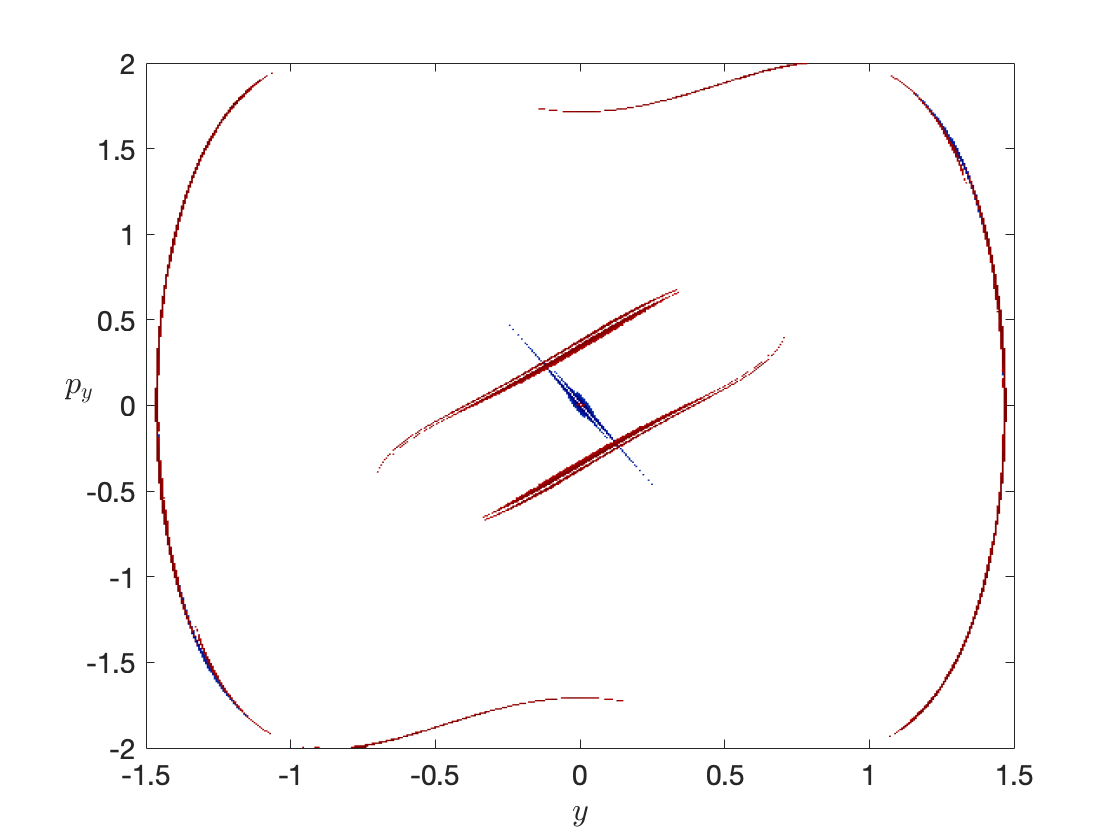}\\
        C)\includegraphics[scale=0.17]{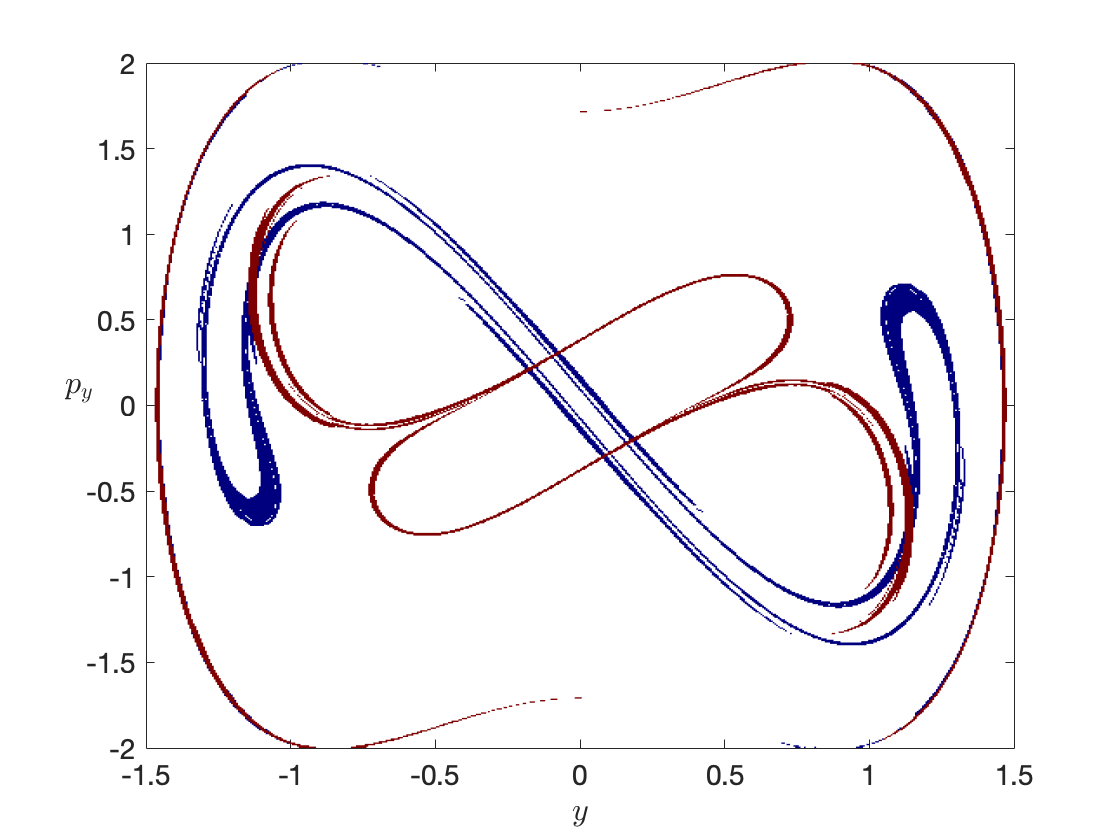}
        D)\includegraphics[scale=0.17]{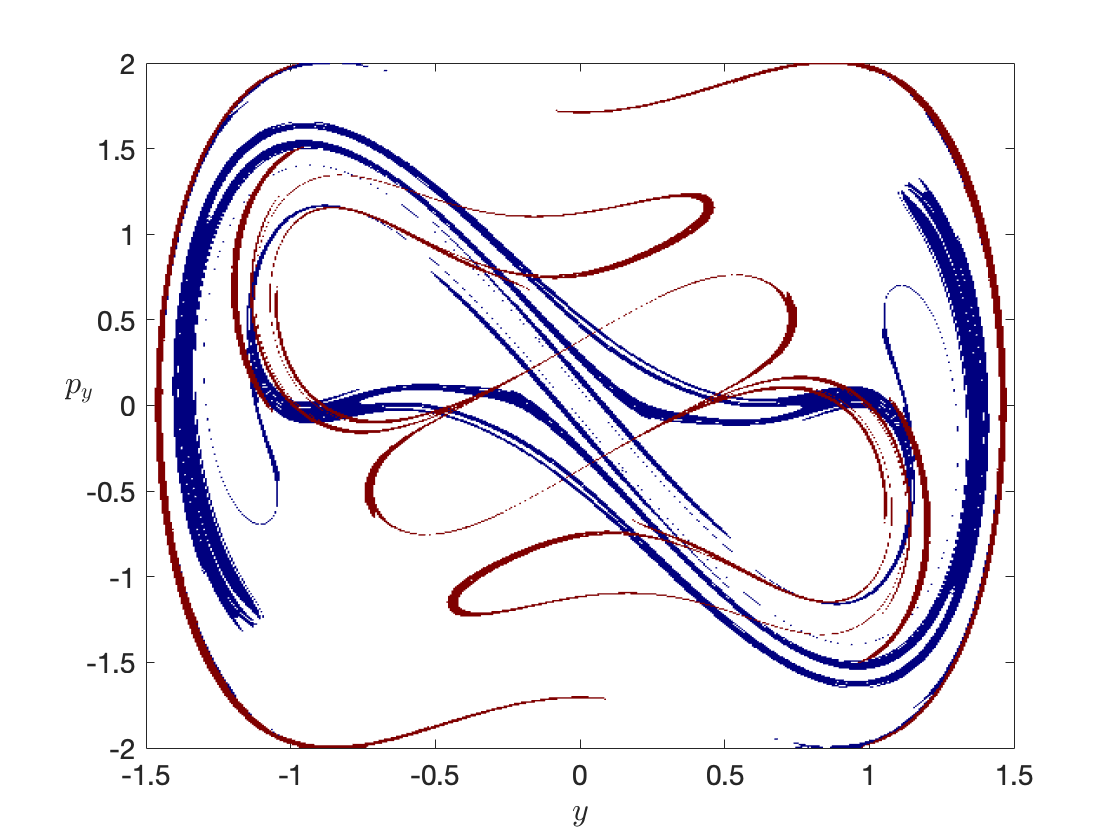}
	\end{center}
	\caption{A)  The unstable (with red color) and  the stable manifolds (with black color) of the top and bottom unstable periodic orbits for $E=0.15$ in the Poincar\'e section $x=1$ with $p_x>0$ (using the classical method).We depict the top (with positive values of $y$) and bottom (with negative values of $y$) periodic orbits by green points. The unstable and the stable invariant manifolds  were computed using the classical method and integrating 160000 initial conditions (for 10 time units) in the direction of the unstable and stable  eigenvector respectively. B-D) Stable and unstable manifolds for $\tau = 2,4,6$ respectively  using the method of LDs.}
	\label{both1015}
\end{figure}

\begin{figure}[htbp]
	\begin{center}
		A)\includegraphics[scale=0.39]{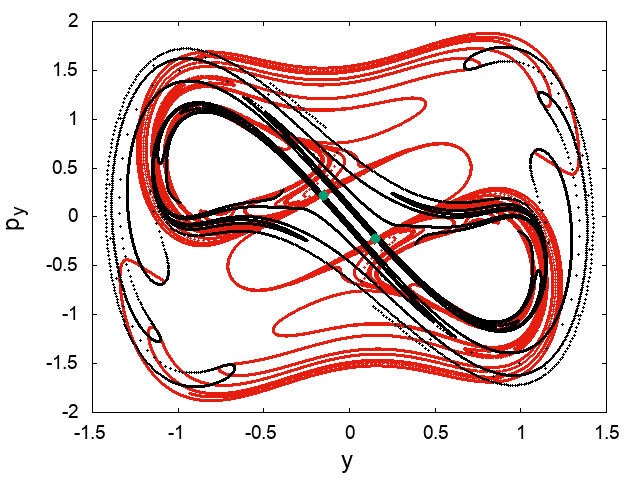}
        B)\includegraphics[scale=0.17]{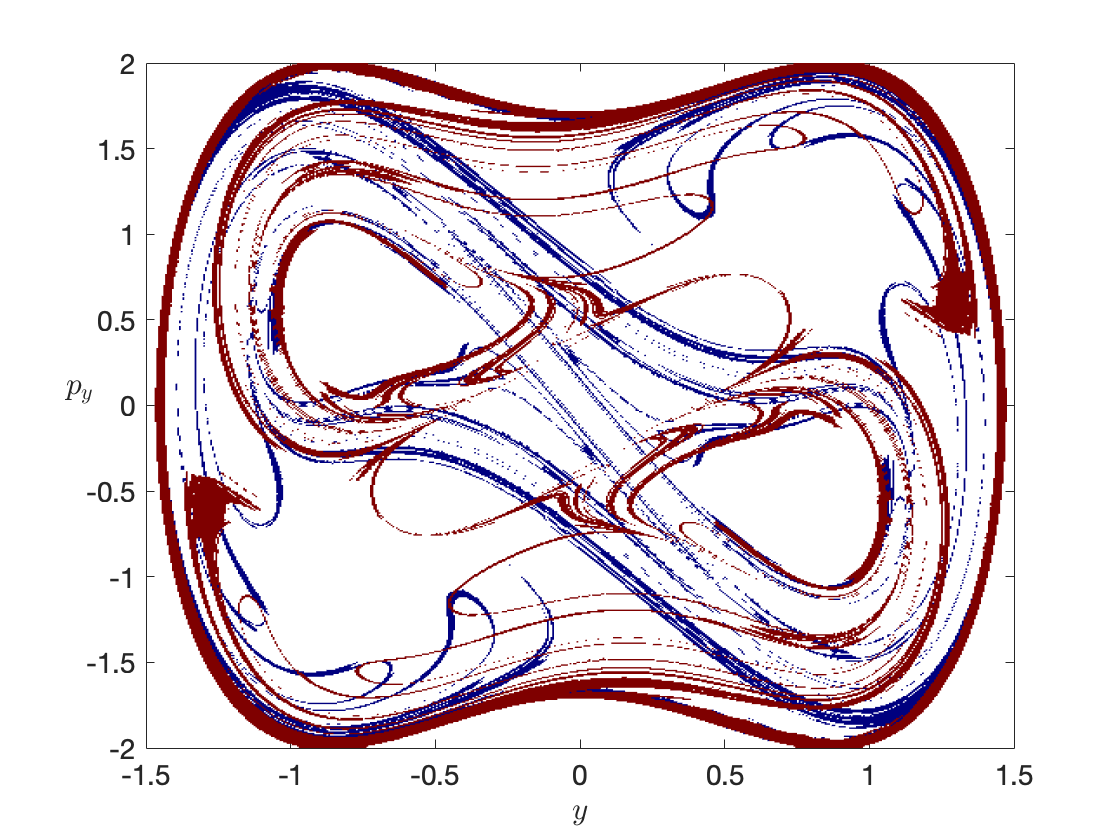}
	\end{center}
	\caption{A) The unstable (with red color) and  the stable manifolds (with black color) of the top and bottom unstable periodic orbits for $E=0.15$ in the Poincar\'e section $x=1$ with $p_x>0$ (using the classical method).We depict the top (with positive values of $y$) and bottom (with negative values of $y$) periodic orbits by green points. The unstable and the stable invariant manifolds  were computed using the classical method and integrating 160000 initial conditions (for 15 time units) in the direction of the unstable and stable  eigenvector respectively. B) Stable and unstable manifolds for $\tau =12$  using the method of LDs.}
	\label{both2015}
\end{figure}

Finally, we compare the ability of the two methods to compute the heteroclinic tangle of the invariant manifolds of the top and bottom unstable periodic orbits. As we observe in panel A of Fig. \ref{both1015} the classical method (for a time integration of 10 time units) detects many heteroclinic and homoclinic points in the central area and on the right and left sides of this tangle and it reveals the complicated structure of this tangle. The heteroclinic tangle that is extracted from the LDs has only heteroclinic and homoclinic intersections in the central area for $\tau=2$ (panel B of Fig. \ref{both1015}) and only other two intersections on left and right side for $\tau=4$ (panel C of Fig. \ref{both1015}). The method of the LDs doesn't reveal the whole tangle for these values of $\tau$. This was something to be expected as we explained before that the LDs, here, doesn't reveal the whole structures for the unstable and stable invariant manifolds for $\tau$ lower than $6$. The heteroclinic tangle that is extracted from the LDs starts to approximate the heteroclinic tangle that is computed from the classical method of Poincar\'e maps for $\tau=6$ (compare the panel A with panel D of Fig. \ref{both1015}). Moreover, if we compute a larger part of this heteroclinic tangle using the classical method (for a larger integration time interval of the initial conditions in the directions of the eigenvectors - This time interval is equal with 15 time units) we see a more complicated structure like this in the panel A of Fig. \ref{both2015}. The LDs approximate this structure for $\tau=12$
(compare panel A with B of Fig. \ref{both2015}).


\begin{figure}[htbp]
	\begin{center}
		A)\includegraphics[scale=0.39]{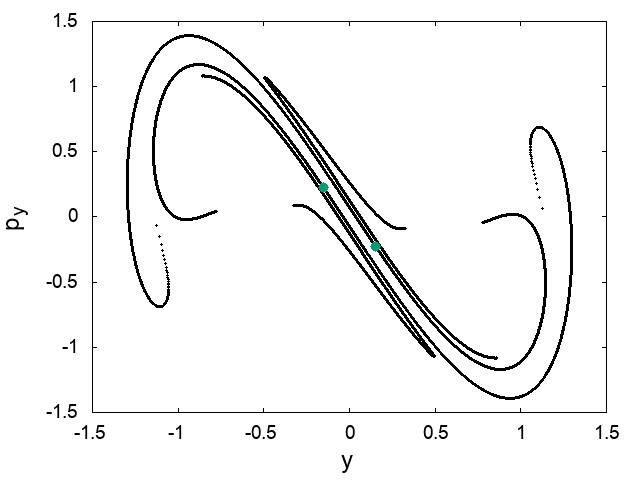}
        B)\includegraphics[scale=0.17]{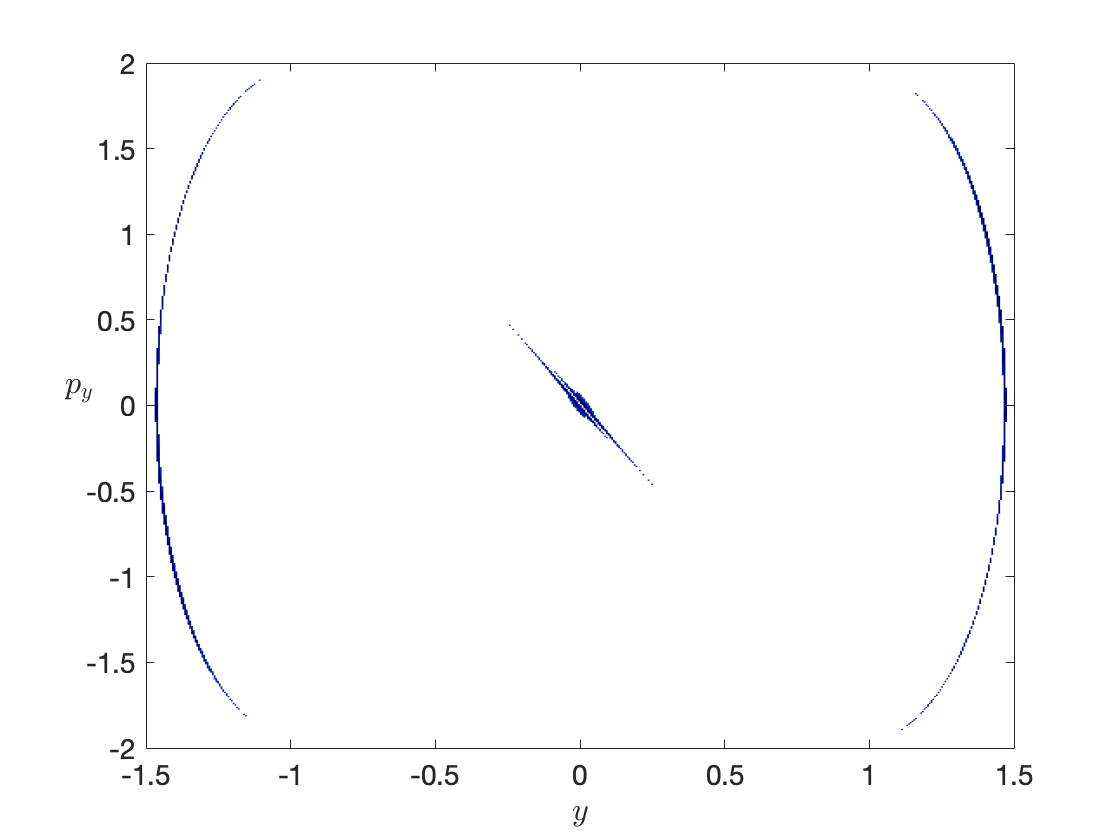}\\
        C)\includegraphics[scale=0.17]{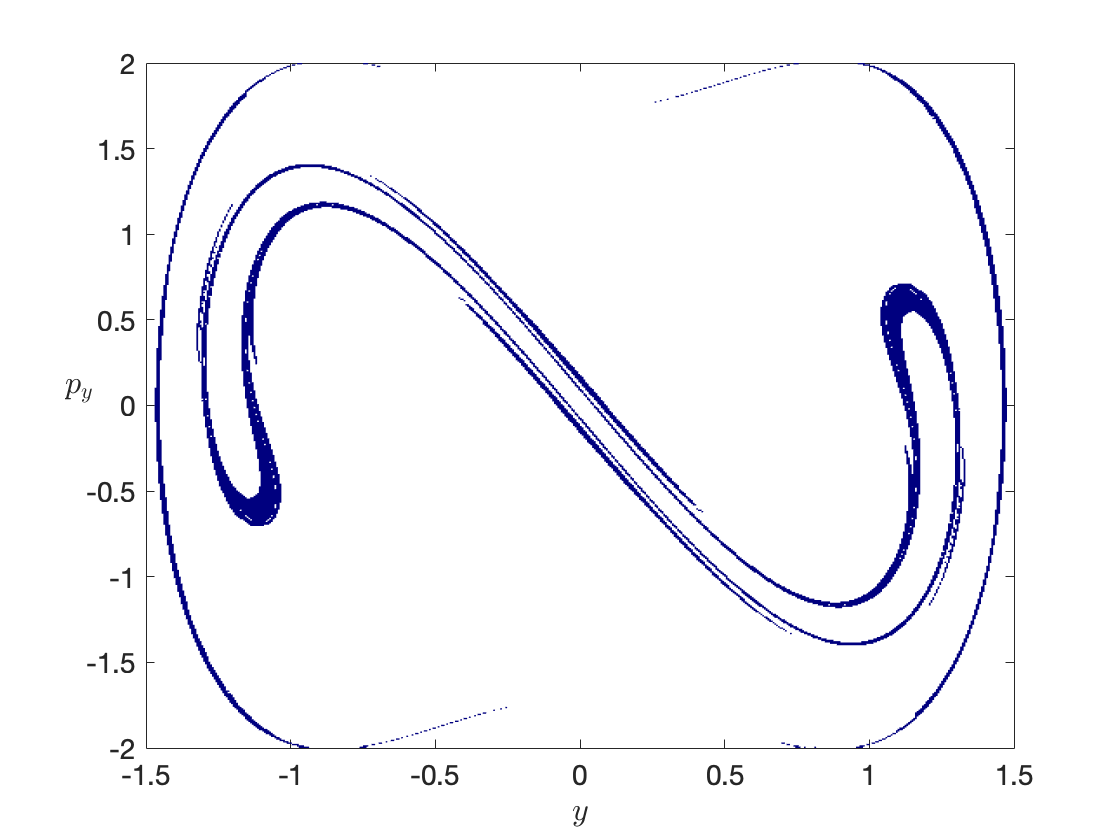}
        D)\includegraphics[scale=0.17]{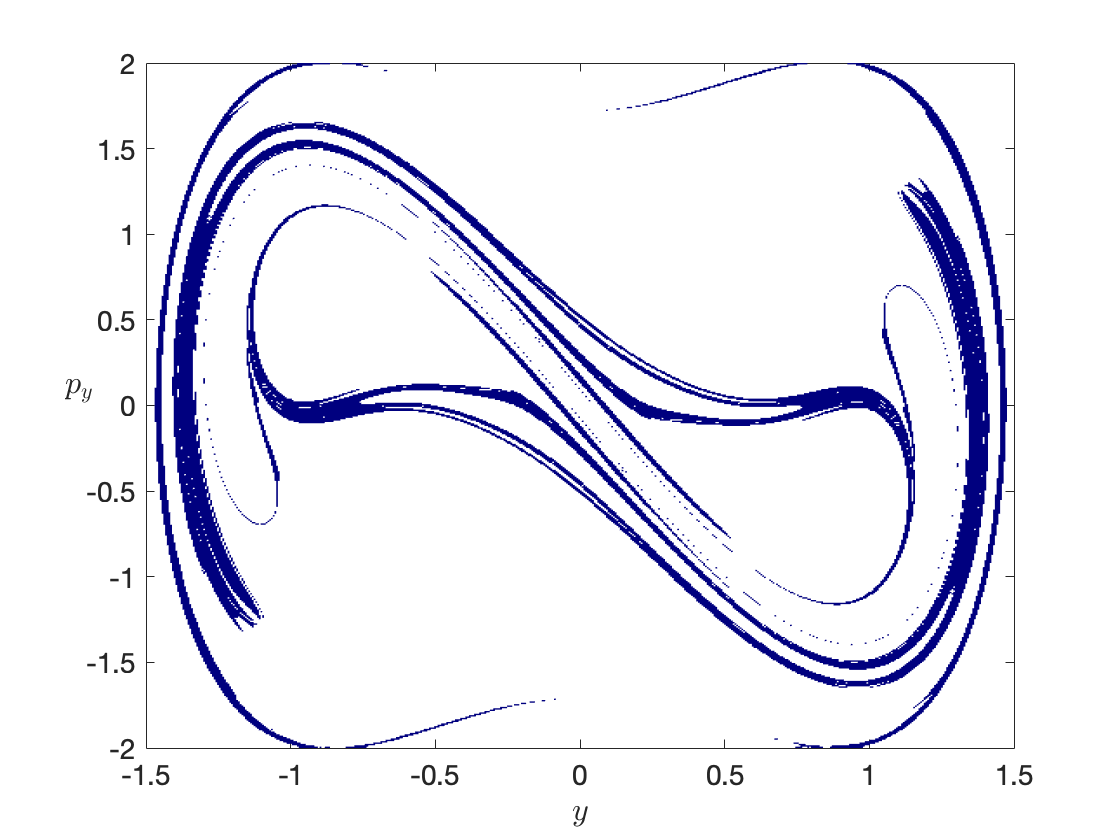}
	\end{center}
	\caption{A) The stable manifolds of the top and bottom unstable periodic orbits  for $E=0.15$ in the Poincar\'e section $x=1$ with $p_x>0$. This was computed using the classical method and integrating 160000 initial conditions (for 10 time units) in the direction of the stable eigenvector. We depict the top (with positive values of $y$) and bottom (with negative values of $y$) periodic orbits by green points.B-D) Stable manifolds for $\tau = 2,4,6$ respectively, using the methods of LDs.}
	\label{st1015}
\end{figure}

\section{Conclusion}\label{summary}
\label{sec:conc}
In this paper, we compared the method of LDs with the traditional method of Poincar\'e maps  in a Hamiltonian system with two degrees of freedom defined by a potential energy surface having two wells and two index-1 saddles previously studied in \cite{collins2011index,Agaoglou2020,katsanikas2020c}. We computed, using these two methods, the invariant manifolds of the unstable periodic orbits that are responsible for the inter-well transport (the transport of the trajectories from the region of the one well to the region of the other well) below and above the energy of the upper index-1 saddle. Our conclusions are the following:

\begin{itemize}
    \item The LDs is a method that is as effective as the method of Poincar\'e maps to extract the homoclinic tangle of the invariant manifolds of the unstable periodic orbits of the lower index-1 saddle that is responsible for the inter-well transport for energy below the energy of the upper index-1 saddle. The advantage of using the LDs method is that there is no need to know the location, the eigenvalues and the eigenvectors of the periodic orbits which you need to know when using the method of Poincar\'e maps.
    \item The LDs is a method that is as effective as the method of Poincar\'e maps to extract the heteroclinic  tangle of the invariant manifolds of the top and bottom unstable periodic orbits that is responsible for the inter-well transport for energy above the energy of the upper index-1 saddle. The advantage of using the LDs method is that there is no need to know the location, the eigenvalues and the eigenvectors of the periodic orbits which you need to know when using the method of Poincar\'e maps.
    \item The integration time interval of the initial conditions that we needed for the method of LDs in order to compute the invariant manifolds of the unstable periodic orbits, for both cases studied in this paper (for energies below and above the energy of the upper index-1 saddle), was 6 and 12 time units which is smaller than the integration time interval that we need for the method of Poincar\'e maps which was 10 and 15.

\end{itemize}
\section*{Acknowledgments}

The authors would like to acknowledge the financial support provided by the EPSRC Grant No. EP/P021123/1.


\clearpage

\appendix

\section{Lagrangian Descriptors}\label{Lag Des}

The method of Lagrangian descriptors (LDs) is a trajectory-based scalar diagnostic that has the capability of revealing the geometrical structures of the phase space. In this paper we have used the p-norm definition of the LDs as it is presented in \cite{lopesino2017}. In particular in this work we have used the value $p=1/2$.

Consider the following dynamical system with time dependence:
\begin{equation}
\dfrac{d\mathbf{x}}{dt} = \mathbf{v}(\mathbf{x},t) \;,\quad \mathbf{x} \in \mathbb{R}^{n} \;,\; t \in \mathbb{R} \;
\label{eq:gtp_dynSys}
\end{equation}
where $\mathbf{v}(\mathbf{x},t) \in C^{r} (r \geq 1)$ in $\mathbf{x}$ and it is continuous in time. Given an initial condition $x_0$ at time $t_0$, take a fixed integration time $\tau>0$ and $p \in (0, 1]$. The method of LDs is as follows:

\begin{equation}
M_p(\mathbf{x}_{0},t_0,\tau) = \sum_{k=1}^{n} \bigg[ \int^{t_0+\tau}_{t_0-\tau}  |v_{k}(\mathbf{x}(t;\mathbf{x}_0),t)|^p \; dt \bigg] = M_p^{(b)}(\mathbf{x}_{0},t_0,\tau)+ M_p^{(f)}(\mathbf{x}_{0},t_0,\tau) \;, 
\label{eq:Mp_function}
\end{equation}

where $M_p^{(b)}$ and $M_p^{(f)}$ its backward and forward integration parts:
\begin{equation}
\begin{split}
M_p^{(b)}(\mathbf{x}_{0},t_0,\tau) &= \sum_{k=1}^{n} \bigg[ \int^{t_0}_{t_0-\tau}  |v_{k}(\mathbf{x}(t;\mathbf{x}_0),t)|^p \; dt \bigg] \;, \\[.2cm]
M_p^{(f)}(\mathbf{x}_{0},t_0,\tau) &= \sum_{k=1}^{n} \bigg[ \int^{t_0+\tau}_{t_0} |v_{k}(\mathbf{x}(t;\mathbf{x}_0),t)|^p \; dt \bigg] \;,
\end{split}
\end{equation}

The forward integration reveals the stable manifolds of our dynamical system. The backward integration reveals the unstable manifolds of our dynamical system. Finally by combining both reveals all the invariant manifolds. Both the stable (blue) and unstable (red) manifolds, as have been presented in the rest of the paper, have been extracted from the gradient of the scalar field generated by LDs.

\begin{figure}[htbp]
	\begin{center}
		A)\includegraphics[scale=0.4]{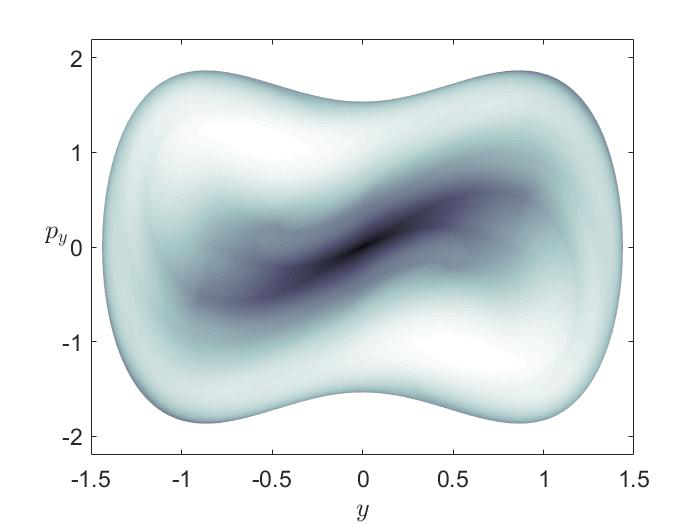}
		B)\includegraphics[scale=0.4]{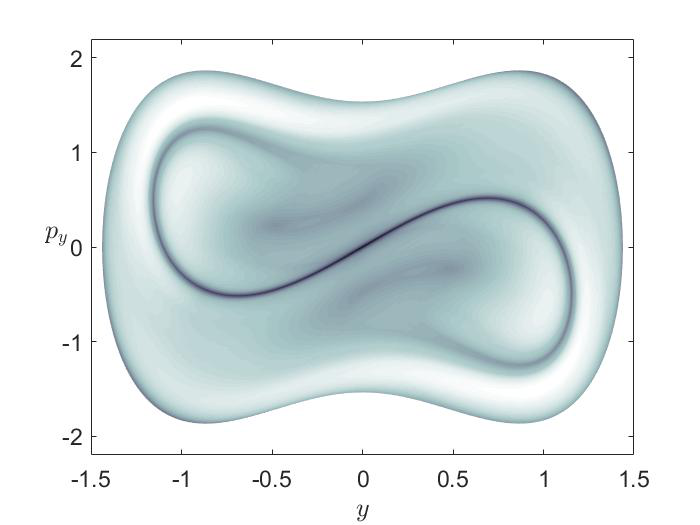}
	    C)\includegraphics[scale=0.4]{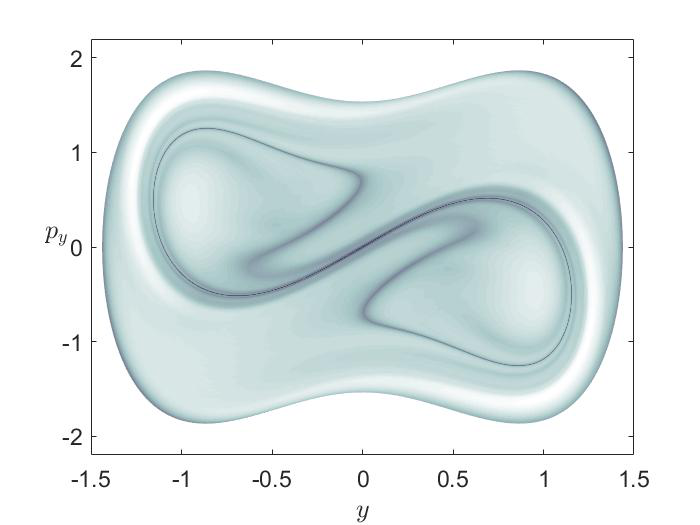}
		D)\includegraphics[scale=0.4]{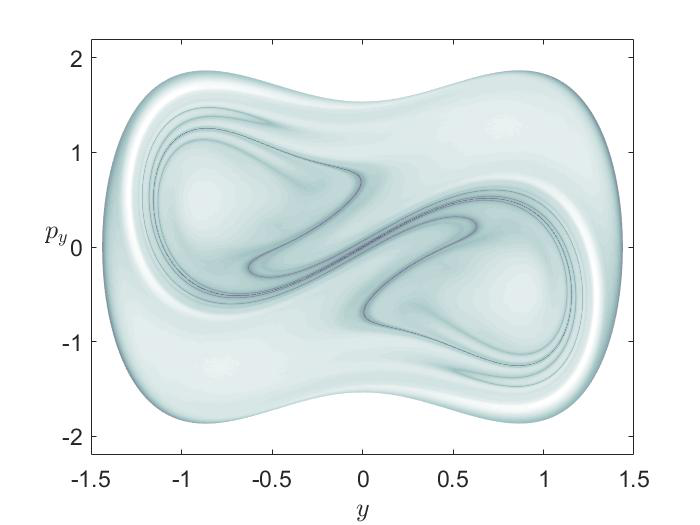}
        \end{center}
	\caption{LDs for $\tau=2,4,6,8$ respectively for energy $H = -0.15$, backward in time.}
	\label{LDs1}
\end{figure}

\begin{figure}[htbp]
	\begin{center}
		A)\includegraphics[scale=0.4]{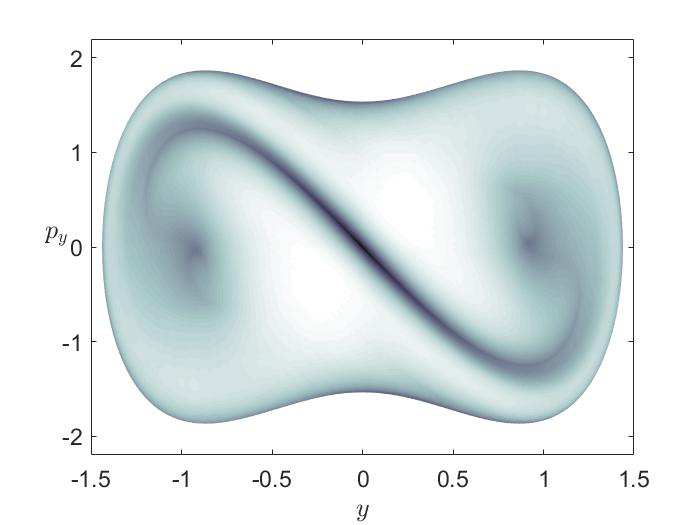}
		B)\includegraphics[scale=0.4]{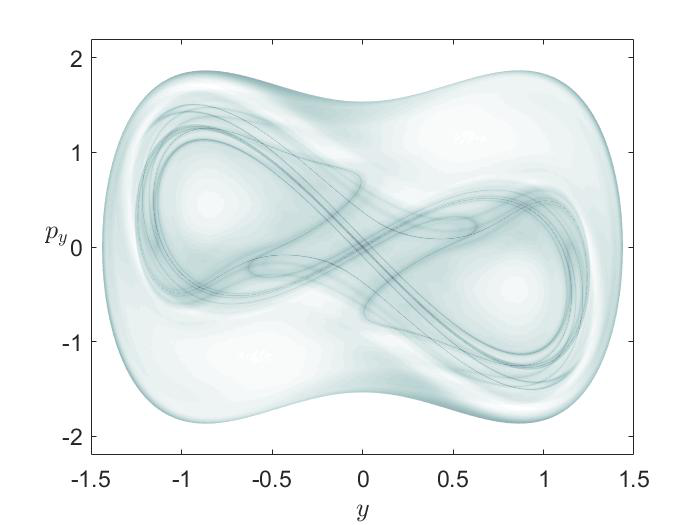}\\
	    C)\includegraphics[scale=0.4]{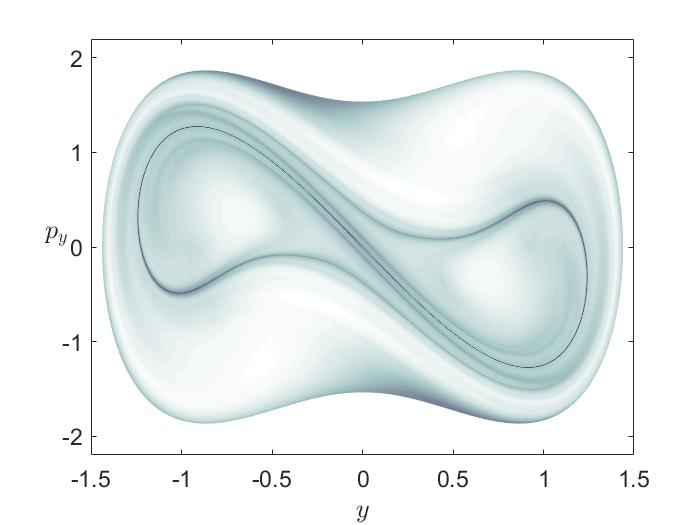}
		D)\includegraphics[scale=0.4]{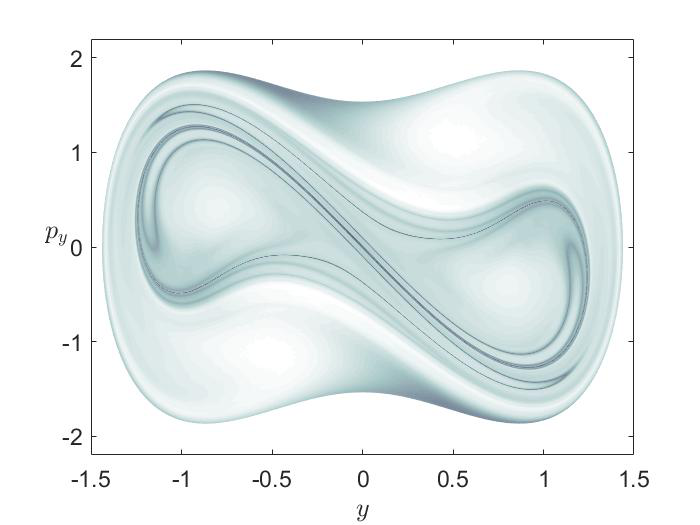}
        \end{center}
	\caption{LDs for $\tau=2,4,6,8$ respectively for energy $H = -0.15$, forward in time.}
	\label{LDs1a}
\end{figure}

\begin{figure}[htbp]
	\begin{center}
		A)\includegraphics[scale=0.4]{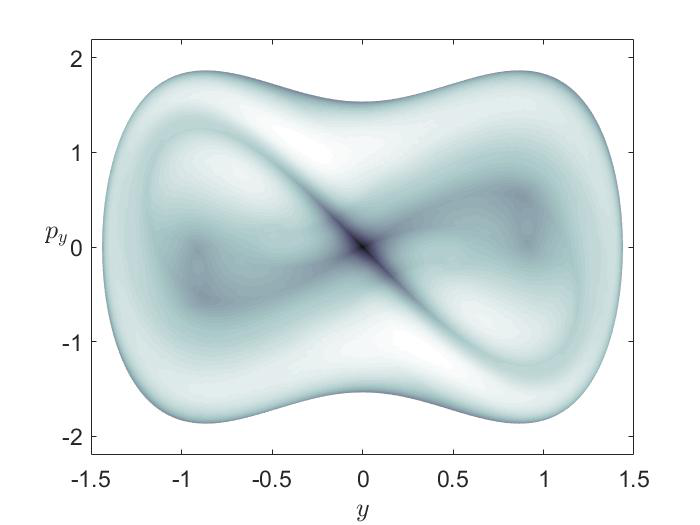}
		B)\includegraphics[scale=0.4]{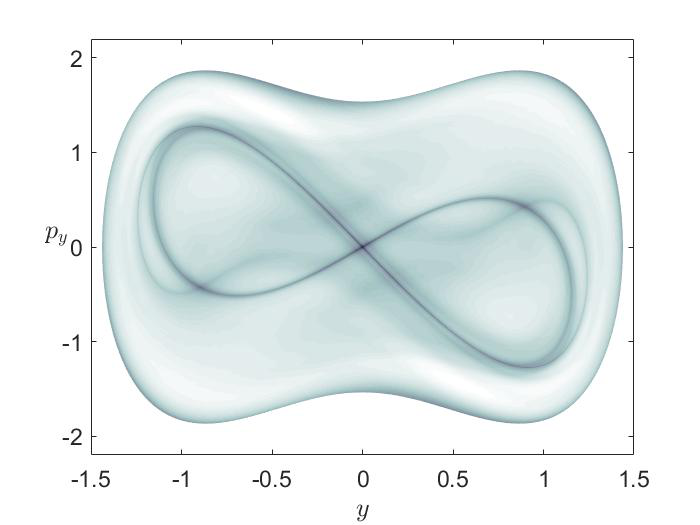}
	    C)\includegraphics[scale=0.4]{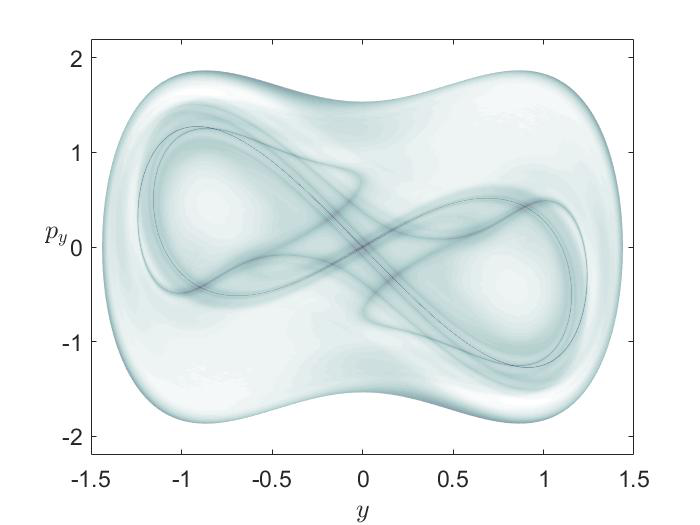}
		D)\includegraphics[scale=0.4]{both_lds_no_title_tau8.png}
        \end{center}
	\caption{LDs for $\tau=2,4,6,8$ respectively for energy $H = -0.15$ backward and forward in time.}
	\label{LDs1b}
\end{figure}

In this appendix we illustrate how the methods of LDs reveals the geometry of invariant manifolds with increasing complexity as the integration time parameter $\tau$ is increased. To show this we have calculate the LDs on the section  \ref{PSOS} for the system with energy $H = -0.15$, that is below that of the index-1 saddle at the origin. For our computations we have used the integration time $\tau = 2, 4, 6,8$. In Fig. \ref{LDs1} we present the LDs backward in time, in Fig. \ref{LDs1a}, forward in time and finally in Fig. \ref{LDs1b} backward and forward in time.

\end{document}